\documentclass[onecolumn,superscriptaddress,preprintnumbers,nofootinbib,amsmath,noshowpacs,aps,prd,floatfix,longbibliography]{revtex4-2}
\usepackage[utf8]{inputenc}
\usepackage{graphicx}
\usepackage{dcolumn}
\usepackage{bm}
\usepackage{float}
\usepackage{tikz}
\usetikzlibrary{quantikz2}
\usepackage{amssymb}
\usepackage{mathtools}
\usepackage{array}
\usepackage{fixmath}
\usepackage{subcaption}
\usepackage{mathrsfs}
\usepackage{xcolor}
\usepackage{bbm}
\usepackage{physics}
\usepackage{multirow}
\usepackage{upgreek}
\usepackage{mathdots}
\usepackage{appendix}
\usepackage{verbatim}
\usepackage[pdftex,colorlinks,linkcolor=black,citecolor=magenta,urlcolor=black]{hyperref}
\usepackage{cleveref}
\usepackage{adjustbox}

\newcommand{\sel}{\textsc{sel}}
\newcommand{\prep}{\textsc{prep}}
\newcommand{\Z}[1]{\mathbbm{Z}_{#1}}
\newcommand{\sfrac}[2]{\sqrt{\frac{#1}{#2}}}

\newcommand{\Set}[1]{\mathcal{S}_{#1}}
\newcommand{\bi}{\boldsymbol{i}}
\newcommand{\bj}{\boldsymbol{j}}

\newcommand{\be}[1]{\boldsymbol{e}_{#1}}

\begin{document}

\title{Resource-optimized fault-tolerant simulation of the Fermi-Hubbard model and high-temperature superconductor models}

\author{Angus Kan}\thanks{akan@psiquantum.com}
\affiliation{PsiQuantum, 700 Hansen Way, Palo Alto, California, 94304, USA}
\affiliation{PsiQuantum, Daresbury, WA4 4FS, UK}

\author{Benjamin C. B. Symons}\thanks{Current affiliation: Department of Physics and Astronomy, University College London, London, WC1E 6BT, UK}
\affiliation{The Hartree Centre, STFC, Sci-Tech Daresbury, Warrington, WA4 4AD, UK}

\date{\today}

\begin{abstract}
Exploring low-cost applications is paramount to creating value in early fault-tolerant quantum computers. Here we optimize both gate and qubit counts of recent algorithms for simulating the Fermi-Hubbard model. We further devise and compile algorithms to simulate established models of cuprate and pnictide high-temperature superconductors, which include beyond-nearest-neighbor hopping terms and multi-orbital interactions that are absent in the Fermi-Hubbard model. We show that simulations of these more realistic models of high-temperature superconductors require only an order of magnitude or so more Toffoli gates than a simulation of the Fermi-Hubbard model. Furthermore, we find plenty classically difficult instances with Toffoli and qubit counts that far lower than commonly considered quantum phase estimation circuits for electronic structure problems in quantum chemistry. We believe our results pave the way towards studying high-temperature superconductors on early fault-tolerant quantum computers.
\end{abstract}

\maketitle


\section{Introduction}

The Fermi-Hubbard (FH) model~\cite{hubbard1964electron} is a longstanding model of interacting electronic systems, which despite its ostensibly simple definition, contains rich physics and is notoriously difficult to simulate classically~\cite{leblanc2015solutions}. The FH model has become a popular subject of fault-tolerant quantum simulation and is widely expected to be one of the earliest physical models to be simulated on a fault-tolerant quantum computer (FTQC) due to its low computational resource requirements. 

In particular, energy estimation algorithms for the FH model on a $L\times L$ square lattice have undergone various optimizations and resource estimations under a fault-tolerant cost model~\cite{wecker2015solving,babbush2018encoding, kivlichan2020improvedfault,lemieux2021resource,campbell2021early,yoshioka2024hunting}.
In~\cite{babbush2018encoding}, Babbush et al. constructed and compiled such an algorithm based on qubitization~\cite{Low2019hamiltonian} down to Clifford + T gates, albeit neglecting the sub-leading order contribution from rotation synthesis. 
In~\cite{kivlichan2020improvedfault}, Kivlichan et al. devised and analyzed an algorithm 
based on second-order Trotter formulas~\cite{suzuki1991general}. They showed that under an extensive error model, i.e., where the error is $O(N)$ where $N$ is the system size~\footnote{They argued that the extensive error model is acceptable in some condensed matter simulations, when, for example, the target observables are robust against a single quasi-particle excitation (see~\cite{babbush2018encoding} for a more detailed discussion).}, the Trotter algorithm can be performed with $O(1)$ T complexity for practically relevant system sizes, compared to the $O(N)$ scaling of the algorithm in~\cite{babbush2018encoding}. Under the more lenient extensive error model, simulations can be performed using orders of magnitude fewer gates, which is favorable to the early FTQC era. More recently, Campbell proposed a new second-order Trotter algorithm with significantly reduced Toffoli count~\cite{campbell2021early}. 

In this work, we optimize and compare the computational resource requirements of the qubitization algorithm~\cite{babbush2018encoding} and Trotter algorithm~\cite{campbell2021early}, under the same extensive error~\footnote{Although we focus on the extensive error model, our optimizations apply to the intensive error model as well.} in~\cite{kivlichan2018quantum,campbell2021early}, i.e., $0.51\% L^2$. We adopt Toffoli~\footnote{Here we assume that 2 T gates can be performed using 1 Toffoli gate with catalyst-assisted circuits from~\cite{Gidney2019efficientmagicstate}.} and logical-qubit count as our cost metrics, which is standard practice in the fault-tolerant algorithm literature, e.g.,~\cite{campbell2021early}.
Despite the expected poorer asymptotic scaling of the qubitization algorithm in this error regime, we still consider it because constant factors are important for small but classically difficult system sizes. 

An often cited motivation behind algorithms to simulate the FH model is to study high-temperature superconductors. It was stated in~\cite{kivlichan2020improvedfault} that the FH model was a candidate model for cuprate high-temperature superconductors, based on various classical simulations in~\cite{leblanc2015solutions}. In~\cite{campbell2021early}, it was stated that the FH model might elucidate the mechanisms behind high-temperature superconductivity. However, more recent numerical evidence~\cite{jiang2019superconductivity,qin2020absence} suggests that in the parameter regime most relevant to cuprate high-temperature superconductors, the FH model does not exhibit superconductivity.

We take the perspective that, to reliably study high-temperature superconductors on a quantum computer, we need to simulate well-established models. We take a step in this direction by designing and compiling Trotter and qubitization algorithms to simulate a single-orbital model for cuprate superconductors~\cite{kim1998Systematics,ANDERSEN1995LDA,delannoy2009low,dalla2012unified,Hirayama2018ab,moree2022ab,lebert2023paramagnon} and a two-orbital model for iron-pnictide superconductors~\cite{Raghu2008minimal,moreo2009properties}. These models are more complex than the FH model, as they contain beyond-nearest-neighbor hopping and multi-orbital interactions, and as such, are harder to classically simulate. To our knowledge, there does not exist any explicit quantum algorithm with resource estimation for simulating such models. This work fills this gap and demonstrates that simulations of the considered more realistic models of high-temperature superconductors only require an increase in Toffoli counts by an order of magnitude or so, when compared to a simulation of the FH model. Furthermore, they can be simulated with fewer computational resources then quantum phase estimation of the electronic structure of many commonly considered molecules~\cite{PhysRevResearch.3.033055,PRXQuantum.2.030305,PhysRevResearch.4.023019,caesura2025fasterquantumchemistrysimulations,low2025fastquantumsimulationelectronic}.

\section{Results}

\noindent\textbf{Trotter schemes for beyond-nearest-neighbor and multi-orbital interactions:} Models of high-temperature superconductors, such as the cuprate and pnictide models considered here, often contain interactions beyond nearest neighbors and between distinct orbitals. Devising Trotter schemes with low Trotter errors and efficient circuit implementations is essential to low-cost simulations of such models. To this end, we design Trotter schemes, for which the Trotter error bounds have constants similarly in magnitude to the constants in the FH Trotter error bounds~\cite{campbell2021early,schubert2023trotter}.

\begin{figure}
    \centering
    \includegraphics[width=\linewidth]{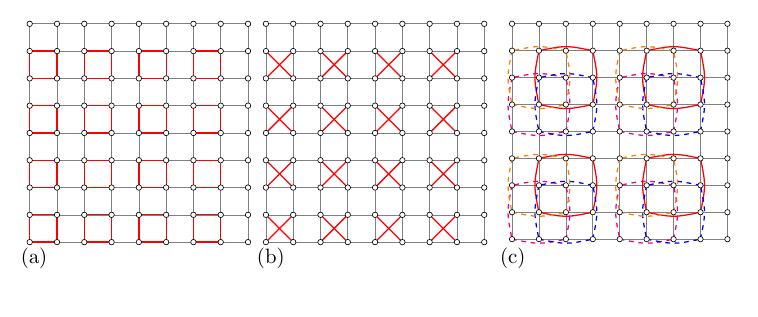}
    \caption{(a) Half of the nn hopping terms in the FH model and cuprate model depicted by the red plaquettes. The other half is obtained by a lattice translation of $\be{x}+\be{y}$. The halves are evolved separately. (b) A quarter of the second-nn hopping terms in the cuprate model depicted by the red crosses. The remaining three quarters are obtained by lattice translations of $\be{x}$, $\be{y}$ and $\be{x}+\be{y}$. Each quarter is evolved separately. (c) Half of the third-nn hopping terms in the cuprate model depicted by both the solid red and dashed lines. The remaining half are obtained by a lattice translation of $2(\be{x}+\be{y})$. The halves are evolved separately.}
    \label{fig:fig1}
\end{figure}

In our schemes, the circuit implementation of the long-range hopping terms share a similar structure to that of the nn hopping in~\cite{campbell2021early}, thereby maximizing the use of HWP. There are three and four types of hopping terms, classified by the magnitudes of their coupling coefficients, in the cuprate and pnictide models, respectively. We implement each type in at most 4 steps (labelled by $l$ in~\eqref{eq:trot}). Upon applications of appropriate fermionic-swap circuits, consisting of only Clifford gates~\cite{kivlichan2018quantum}, each segment can be localized to nn hopping terms acting on either non-overlapping (i) plaquettes, as depicted in figure~\ref{fig:fig1}(a), or (ii) links. For example, the second-nn hopping terms in figure~\ref{fig:fig1}(b) will be mapped to case (ii), if we swap to every even site on every even row, assuming a zero-based numbering, with the site directly above. Case (i) can be implemented using the aforementioned circuit from~\cite{campbell2021early}. The circuit implementation of case (ii) is even simpler. The hopping term on each link is of the form $e^{i\theta (XX+YY)}$, which can be diagonalized to a layer of two same-angle $R_z$ gates using Clifford gates~\cite{kivlichan2020improvedfault,Wang2021resourceoptimized}. As such, case (ii) can be implemented using HWP, up to Clifford gates. The on-site Coulomb term in the cuprate model and the intra-orbital Coulomb term in the pnictide model are implemented the same way as the Coulomb term in the Fermi-Hubbard model; the implementation of the inter-orbital Coulomb term in the pnictide model is almost the same, except fermionic swaps are used to localize them to two-body $ZZ$ operators, which are then implemented using Clifford gates and HWP. We defer further details of the Trotter schemes for the three models to Supplementary Note 2.

\noindent\textbf{Space-time trade-off in Hamming-weight phasing:} In the standard, \emph{baseline} HWP~\cite{Gidney2018halvingcostof,nam2019low,campbell2021early}, the Hamming weight of the to-be-rotated $M$-qubit state is first computed into an ancilla register of size $\lfloor \log_2(M) \rfloor + 1$. Then, $\lfloor \log_2(M) \rfloor + 1$ $R_z$ rotations $\bigotimes_{i=0}^{\lfloor \log_2(M) \rfloor} R_z(2^i \theta)$ are applied to the ancilla register, before the Hamming weight is uncomputed. It was shown in~\cite{muller1975bounds} that the Hamming weight of a binary register can be computed using $M - w(M)$ half- and full-adders, where $w(M)$ is the number of 1s in the binary representation of $M$. Briefly, starting from the lowest bit of the binary register, we perform a chain of summations, using full adders as much as possible; each sum bit becomes an input to the next adder until the last adder, and the carry bits become inputs to the next chain of summations. This carries on recursively until there are no more bits to add. A half- or full-adder can be implemented, up to Clifford gates, with a Toffoli gate, and can be uncomputed via a measurement-feed-forward Clifford circuit~\cite{Gidney2018halvingcostof}. (See Supplementary Figure 2 for the half- and full-adder circuits, and an example of an 8-bit Hamming-weight computation circuit.) This means that the computation and un-computation of the Hamming-weight function cost $M - w(M)$ Toffoli gates. 

We introduce a gate-optimized version of HWP, which we call \emph{catalyzed} HWP. In catalyzed HWP, $\bigotimes_{i=0}^{\lfloor \log_2(M) \rfloor} R_z(2^i \theta)$ is implemented via a generalized phase gradient operation (see equation 168 in~\cite{wang2024option}; also shown in Supplementary Figure 3). Briefly, the operation applies an adder-like circuit, consisting of $\lfloor \log_2(M)\rfloor+1$ Toffoli gates and a single $R_z$ gate, to a catalyst state of the form $\bigotimes_{i=0}^{\lfloor \log_2(M) \rfloor} R_z(2^i \theta)\ket{+...+}$ and the to-be-rotated target state. Since the catalyst state comes out of the operation unscathed and is thus reusable, its preparation costs are amortized over the course of a simulation. Note that in general, multiple catalyst states are needed to simulate $H$ because each $H_l$ in~\eqref{eq:trot} is associated with a different $\theta$-value and different $\theta$-values require different catalyst states. To summarize, an application of baseline HWP requires $M-w(M)$ Toffoli gates and $\lfloor \log_2(M)\rfloor + 1$ $R_z$ rotations, whereas an application of catalyzed HWP requires $M+\lfloor \log_2(M)\rfloor -w(M)+1$ Toffoli gates and only one $R_z$, neglecting, for the moment, the synthesis cost of the catalyst state, which we will take into consideration in our resource estimation, and using only logarithmically more ancilla qubits. This trade-off is worthwhile because Toffoli gates are cheaper than $R_z$ gates on a FTQC~\cite{Gidney2019efficientmagicstate,Kliuchnikov2023shorterquantum}.

\begin{figure}
    \centering
    \includegraphics[width=\linewidth]{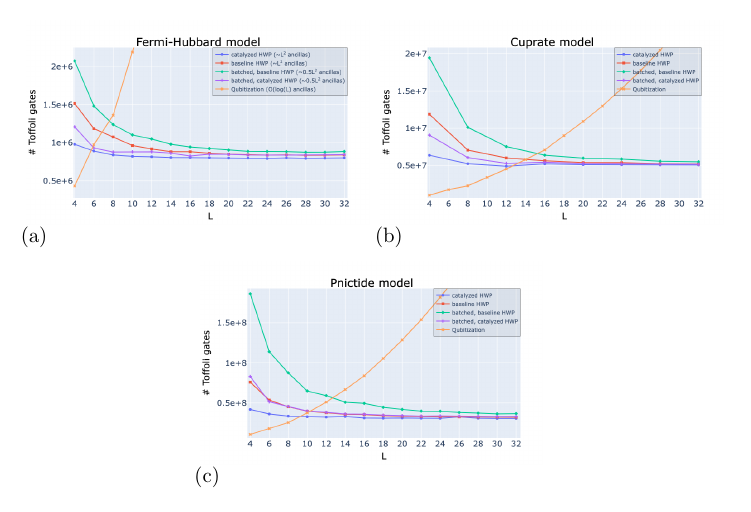}
    \caption{\textbf{Toffoli counts for (a) Fermi-Hubbard simulations with $u=8$ and $t=1$, (b) cuprate simulations with $u=8, t=1, t'=0.3$ and $t''=0.2$, and (c) pnictide simulations with $u=8, t_1=1, t_2=1.3, t_3 = 0.85$, and $t_4=0.85$.} The explicit Toffoli counts in these plots can be found in Supplementary Note 1 and Supplementary Note 2. The logical qubit counts of these simulations are shown in figure~\ref{fig:fig3}. Notably, compared to the FH model, the cuprate and pnictide models require only about one order and less than two orders of magnitude more Toffoli gates to simulate. Note that the resource estimation shown here is carried out for an extensive target error of $0.51\% L^2$.}
    \label{fig:fig2}
\end{figure}

In the qubit-limited, early FTQC era, it is desirable to limit the number of ancilla qubits. In both baseline and catalyzed HWP, we can perform a layer of same-angle $R_z$ rotations in batches, instead of all at once, which reduces the ancilla-qubit count but sacrifices some gate efficiency. In~\cite{campbell2021early}, the number of same-angle $R_z$ rotations applied per batch was limited to $0.5L^2$ so that $\sim 0.5L^2$ ancilla qubits would be required. We apply baseline and catalyzed HWP, without batching and in batches of $0.5L^2$ rotations, to implement the same-angle $R_z$ gates in the simulations of the FH, cuprate, and pnictide models, and plot their quantum resource estimates in figure~\ref{fig:fig2}. For the FH model, we set $u=8$ and $t=1$, which was identified in~\cite{leblanc2015solutions} as the parameter setting with the largest uncertainty in classical simulations. For the cuprate model, we set $u=8, t=1, t'=0.3$ and $t''=0.2$, in line with parameter choices for cuprate simulations in the literature~\cite{delannoy2009low,dalla2012unified,Hirayama2018ab,moree2022ab,lebert2023paramagnon}. For the pnictide model, we set $t_1=1, t_2=1.3, t_3 = 0.85, t_4=0.85$, as per~\cite{Raghu2008minimal}, and $u=8$, motivated by the classical hardness of FH simulations at $u/t=8$. We refer readers to Supplementary Note 2 for more details on the resource estimation pipeline.

We compare the four different Trotter implementations and highlight the key findings:
\begin{itemize}
    \item Our Toffoli count estimates for the FH, cuprate, and pnictide simulations plateau around $8\times 10^5, 5\times 10^6$, and $3\times 10^7$, which are in line with the $O(1)$ T complexity described in~\cite{kivlichan2020improvedfault}. These numbers are orders of magnitude smaller than typical resource estimates for QPE circuits for moderately sized electronic structure problems~\cite{PRXQuantum.2.030305,caesura2025fasterquantumchemistrysimulations,low2025fastquantumsimulationelectronic,PhysRevResearch.3.033055,PRXQuantum.2.030305,PhysRevResearch.4.023019,caesura2025fasterquantumchemistrysimulations,low2025fastquantumsimulationelectronic}, which often require $\gtrsim 10^9$ Toffoli gates and $\gtrsim 10^3$ logical qubits, thereby enriching the early FTQC application landscape. The sub-million Toffoli counts for FH simulations, especially when the system size is addressable with $<10^3$ logical qubits, make them particularly attractive and suitable to become one of the earliest scientific application of FTQCs.
    \item The implementations based on catalyzed HWP, batched or not, are always more gate-efficient than those based on baseline HWP. In particular, for the smallest classically difficult square system size, i.e., $8 \times 8$~\cite{leblanc2015solutions,liu2025accuratesimulationhubbardmodel}, a reduction of $>20\%$ in Toffoli count is achieved for a FH simulation. Note that the estimates for (batched,) baseline HWP in figure~\ref{fig:fig2}(a) are obtained from our re-compilation of the algorithm in~\cite{campbell2021early}. The gate-reductions become smaller at larger system sizes because of the lenient, extensive error model; the improvements in gate-efficiency will be more pronounced for applications with smaller error tolerance.
    \item Simulations using batched, catalyzed HWP can achieve roughly the same or even better gate-efficiency than those using baseline HWP, while employing only a fraction -- roughly $1/2, 1/4$, and $1/8$ for FH, cuprate, and pnictide simulations respectively -- of the ancilla qubits required for baseline HWP.
\end{itemize}

\noindent\textbf{Qubitization vs. Trotter comparison:} We compile qubitization algorithms to simulate the FH, cuprate, and pnictide models, and perform resource estimation for them. Even though their Toffoli complexity, with respect to the system size, is worse than their Trotter counterparts, we want to find out whether there is a range of system sizes where their Toffoli counts are better than or close to their Trotter counterparts. We optimize the qubitization algorithms by applying the same chemical potential shift, i.e., $n\mapsto n-1/2$, from~\cite{campbell2021early}, which we have used in our Trotter algorithms. This shift reduces the number of Pauli terms in the Hamiltonian after Jordan-Wigner transformtion; thus, it reduces the induced 1-norm $\lambda$ and in turn, the Toffoli counts of qubitization algorithms, which we plot in figure~\ref{fig:fig2}. However, the Toffoli complexity still scales as $L^2$, with sub-leading additive costs that scale logarithmically in $L$, as does the algorithm in~\cite{babbush2018encoding}. We find the crossover points in $L$ the linear dimension, where qubitization becomes more expensive in Toffoli count than all Trotter implementions, to be $L=8,16$, and $14$ for FH, cuprate, and pnictide simulations, respectively. It is important to point out that qubitization algorithms have a lower ancilla count than Trotter algorithms -- $O(\log(L))$ vs. $O(L^2)$ -- due to the spatial overhead of HWP. A potential consequence is that in the qubit-limited, early FTQC era, qubitization simulations of smaller system sizes might still be favorable despite their worse Toffoli count, especially as magic states become cheaper~\cite{gidney2024magicstatecultivationgrowing,wan2024constanttimemagicstatedistillation}.

\section{Discussion}

In this work, we devise resource-optimized energy estimation algorithms for the Fermi-Hubbard model, a single-orbital model of cuprate superconductors, and a two-orbital model of iron pnictide superconductors. Our algorithms are based on the second-order Trotter formula and qubitization. We introduce a technique called catalyzed Hamming-weight phasing, which is particularly effective at reducing the rotation synthesis costs of our Trotter algorithms. Furthermore, comparing our Trotter and qubitization algorithms, we locate precisely the system sizes where qubitization could be more advantageous, despite its worse gate complexity under the extensive error model. Our resource estimation indicates the simulations of the considered cuprate and pnictide models are mildly costlier than the FH model by one order of magnitude or so in terms of Toffoli counts.

When paired with a preparation protocol for an initial state that has a good overlap with the ground state, which is beyond the scope of this work, our algorithms can be used to perform ground state energy estimation (GSEE)~\cite{PhysRevLett.121.010501,lemieux2021resource,wecker2015solving}. Typically, the overall cost of GSEE is $O(1/\gamma^b)$ times the QPE cost, where $\gamma$ is the overlap and $b$ depends on the chosen method. Research for initial state preparation has been explored in, e.g.,~\cite{lemieux2021resource}, which applies an adiabatic algorithm to improve matrix-product-state (MPS) estimates in the context of the FH model, and more recently in~\cite{PRXQuantum.5.040339,berry2024rapidinitialstatepreparation}, which propose more optimized algorithms based on MPS. The resource estimates from~\cite{PRXQuantum.5.040339,berry2024rapidinitialstatepreparation} suggest that the costs of preparing high-quality initial states are similar to or less than the QPE costs, in the context of electronic structure GSEE; whether similar conclusions hold for fermionic lattice simulations are worth investigating by future work. In particular, MPS simulations of the two-dimensional Hubbard model~\cite{liu2025accuratesimulationhubbardmodel} could require higher bond dimensions than those of the electronic structure of oft considered molecules, e.g., FeMoco. Since the algorithms in~\cite{PRXQuantum.5.040339,berry2024rapidinitialstatepreparation} scale with bond dimension, the state preparation cost of Hubbard models could exceed that of FeMoco, which would be more expensive than the QPE cost of Hubbard models. This could potentially be alleviated by considering alternative tensor-network states such as fermionic projected entangled pair states (PEPS) which can achieve similar accuracies to MPS at a much lower bond dimension~\cite{liu2025accuratesimulationhubbardmodel}. Alternatively, one could consider Lindbladian-based ground-state preparation methods~\cite{PhysRevResearch.6.033147,smid2025polynomialtimequantumgibbs}. For such methods to be efficient, one needs to show that the simulated system's mixing time scales polynomially with the system size. It has been proven for certain parameter regimes, the Hubbard model exhibits polynomial mixing time~\cite{smid2025polynomialtimequantumgibbs}; however, crucially, for the interesting, intermediate $U/t$-regime, only numerical evidences for fast mixing times exist for small lattices~\cite{smid2025polynomialtimequantumgibbs}. In~\cite{zhan2025rapidquantumgroundstate}, one could find positive numerical results for larger lattices, though not of the Hubbard model. In the future, the practicality of any initial state preparation method -- whether it be tensor-network or Lindbladian method -- will need to be evaluated via rigorous resource estimation, just as the QPE part has been over the years.

Comparisons between classical and quantum GSEE of the FH model are of interest from the quantum -- as well as the wider -- community; see~\cite{yoshioka2024hunting,Lee2023} for quantitative comparisons. The conclusions drawn from any comparison will necessarily depend on the state of both classical and quantum simulation, which are constantly improving, as evidenced by this work. As it currently stands, when considering the FH model on square lattices, $8\times 8$ is the largest system size that can be reliably simulated using the state-of-the-art classical algorithms based on MPS~\cite{leblanc2015solutions,liu2025accuratesimulationhubbardmodel}. Simulations of more realistic and complex Hubbard-like models, with beyond-nearest-neighbor hopping and multi-orbital interactions, are likely limited to smaller lattices, although more concrete conclusions are better drawn after more rigorous studies in the future. Since the works of~\cite{yoshioka2024hunting,Lee2023} and the completion of this work, recent advances in classical tensor-network simulations based on PEPS have been reported in~\cite{liu2025accuratesimulationhubbardmodel}; the results therein show ground state energy estimates that are similar but not always better than MPS results at small system sizes and that are, to some extent, in qualitative agreement with theory at system sizes beyond the reach of MPS simulations. We expect larger system sizes to be within the reach of classical tensor-network simulations in the future, and that such developments in classical tensor-network simulations will lead to improvements in quantum initial state preparation protocols, just as classical MPS simulations did. We remark that there could potentially be scientific value in performing quantum simulations even at system sizes that are reachable by classical tensor-network simulations, because the latter are heuristic and tensor-network, e.g., MPS and PEPS, estimates could be improved with more rigorous performance guarantees on a quantum computer using, e.g., algorithms in~\cite{lemieux2021resource}. We believe any further nuanced classical-versus-quantum comparisons will continue to garner interests and are best addressed outside of this work.

We list several other interesting directions to be explored in the future below:
\begin{itemize}
    \item There are other problems beyond GSEE, which could play to the strengths of quantum computers. For example, the estimation of dynamical observables, e.g., dynamical correlations~\cite{wecker2015solving}, which involves time-evolution of a quantum state that can be performed efficiently, using, e.g., our Trotter and qubitization circuits, on a quantum computer but is, to our knowledge, exponentially expensive on a classical computer. Gleaning ground-state observables, other than the energy, are relevant in physics, but can be more expensive than GSEE; see~\cite{Steudtner2023faulttolerant} for an example in the context of molecular simulations. Developing concrete quantum algorithms to estimate these observables are paramount to expanding the utility of quantum computers.
    \item We have assumed worst-case Trotter error bounds in our work. However, it is worth investigating the cost reductions made possible when average-case empirical Trotter errors, which are often smaller than worst-case bounds~\cite{schubert2023trotter}, are assumed.
    \item Hubbard-like lattice models are low-energy effective models of real-world strongly-correlated materials~\cite{hubbard1964electron}; they can be derived by keeping very few salient microscopic interactions to make theoretical and numerical studies tractable. The more microscopic interactions a model includes, the more realistic the model is, but it becomes costlier to simulate. It could be interesting to better understand this trade-off by obtaining resource estimates for quantum simulations of cuprate~\cite{ANDERSEN1995LDA,moree2022ab} and pnictide models~\cite{PhysRevLett.101.087004} that contain interactions between more orbitals than the models considered in this work.
    \item Construct simulations of Hubbard-like models interacting with a classical field to probe photoexcitation effects~\cite{watzenbock2022photoexcitations}, using time-dependent simulation algorithms~\cite{low2019hamiltoniansimulationinteractionpicture,prep}.
    \item Perform a more detailed fault-tolerant cost analysis under computational models that incorporate the cost of Clifford gates, e.g., the active volume model~\cite{litinski2022activevolumearchitectureefficient}; costing Clifford gates will become more important as magic states become cheaper~\cite{gidney2024magicstatecultivationgrowing,wan2024constanttimemagicstatedistillation}.
\end{itemize}

\section{Methods}

\noindent \textbf{Target models:} Six decades from its inception~\cite{hubbard1964electron}, the FH model remains intensely studied because it contains very few physical degrees of freedom, yet it captures salient features of strongly correlated electronic systems. Despite the attention it garners, it has been exactly solved only in one dimension~\cite{metzner1989correlated} and infinite dimensions~\cite{lieb1968absence}. In two and three dimensions, which are more relevant to realistic materials, exact solutions remain elusive and numerical solutions are intractable for even moderate system sizes. Here we focus on the two-dimensional case. 

We consider a fermionic Hamiltonian $H$ consisting of a hopping part $H_h$ and on-site Coulomb contribution $H_c$. $H_h$ is given by
\begin{equation}
    H_h = \sum_{\substack{(\bi,\bj) \in \Set{1}, \\ (f,f') \in \Set{2}}} T_{(\bi,f),(\bj,f')} a^{\dag}_{\bi,f} a_{\bj,f'},
\end{equation}
where $\bi, \bj$ label lattice sites and $f,f'$ label internal degrees of freedom, e.g., spins and orbitals, $T_{(\bi,f),(\bj,f')}$ is a Hermitian matrix that stores the coupling coefficients, $a^{(\dag)}$ is the annihilation (creation) operator, and the content of sets $\Set{1}$ and $\Set{2}$ depend on the model considered. $H_c$ is given by
\begin{equation}
    H_c = \sum_{\substack{\bi \in \Omega, \\ (f,f')\in \Set{3}}} U_{f,f'} n_{\bi, f}n_{\bi, f'},
\end{equation}
where $n=a^\dag a$ is the number operator, $\Omega$ is the set of all lattice sites, the matrix $U_{f,f'}$ stores the on-site coefficients, and $\Set{3}$ is a model-dependent set. We consider an $L \times L$ lattice with an even $L$ and periodic boundary conditions such that $\Omega = (\Z{L})^2 = \{0,1,...,L-1\}^2$. Hereafter, we denote the unit vectors $(1,0),(0,1) \in \Omega$ as $\be{x},\be{y}$, respectively.

For the FH model, $\Set{1}$ contains all nearest-neighbor (nn) pairs, $f$ and $f'$ label the spin, $\Set{2} = \{(\uparrow,\uparrow), (\downarrow, \downarrow) \}$, and $\Set{3} = \{(\uparrow,\downarrow), (\downarrow, \uparrow) \}$. The non-zero elements of $T_{(\bi,f),(\bj,f')}$ and $U_{f,f'}$ are $t$ and $u$, respectively. For the single-orbital cuprate model~\cite{kim1998Systematics,ANDERSEN1995LDA,delannoy2009low,dalla2012unified,Hirayama2018ab,moree2022ab,lebert2023paramagnon}, $\Set{1}$ additionally includes second- and third-nn, and $T_{(\bi,f),(\bj,f')}$ has three types of distinctly valued, non-zero elements $t$, $t'$, and $t''$, which are the coupling between first-, second-, and third-nearest neighbors, respectively.

For the two-orbital pnictide model~\cite{Raghu2008minimal,moreo2009properties}, $\Set{1}$ contains first- and second-nn, $f = (\sigma, d)$, where $\sigma$ denotes the spin and $d \in \{ x,y \}$ labels the orbital, and $\Set{2} = \{ ((\sigma, d),(\sigma, d')) | \sigma\in \{ \uparrow, \downarrow\},\: (d,d') \in \{ (x,x),(y,y),(x,y),(y,x) \}  \}$. $T_{(\bi,f),(\bj,f')}$ has five types of distinctly valued, non-zero elements: (i) $t_1$ couples nearest neighbors in the $\be{y}$-direction with orbitals $(d,d') = (x,x)$ and those in the $\be{x}$-direction with $(d,d') = (y,y)$, (ii) $t_2$ couples nearest neighbors in the $\be{x}$-direction with $(d,d') = (x,x)$ and those in the $\be{y}$-direction with $(d,d') = (y,y)$, (iii) $t_3$ couples second-nearest neighbors with the same orbital, (iv) $t_4$ and (v) $-t_4$ couple second-nearest neighbors separated by $\be{x} + \be{y}$ and $\be{x} - \be{y}$, respectively, with different orbitals. $U_{f,f'}$ contains two distinctly valued, non-zero elements: (i) the intra-orbital Coulomb strength $u$ for $(f,f') \in \{((\uparrow,d),(\downarrow,d)) | d\in \{x,y\} \}$ and (ii) inter-orbital Coulomb strength $v$ for $(f,f') \in \{((\sigma,x),(\sigma',y)) | \sigma,\sigma' \in \{\uparrow,\downarrow\} \}$. Here we do not consider other weaker types of on-site interaction, such as Hund's rule coupling and pair hopping that are over an order of magnitude weaker than Coulomb interactions~\cite{Raghu2008minimal,moreo2009properties}.

\noindent \textbf{Related work:} Energy estimation algorithms for the FH model have been studied extensively~\cite{wecker2015solving,babbush2018encoding, kivlichan2020improvedfault,lemieux2021resource,campbell2021early,yoshioka2024hunting}. Here we briefly review the state-of-the-art qubitization~\cite{babbush2018encoding} and Trotter algorithm~\cite{campbell2021early}, which form the basis of our algorithms.

Qubitization~\cite{Low2019hamiltonian} assumes an input model, where the target Hamiltonian $H$ is expressed as a linear combination of unitaries (LCU), i.e.,
\begin{equation}
    H = \sum_{i=1}^{M} c_i P_i, \: c_i \in \mathbbm{R}, \: c_i \geq 0,
\end{equation}
where $P_i$'s are unitary and self-inverse, e.g., Pauli operators. Then, $H$ is encoded using the following oracles:
\begin{gather}
    \prep \ket{0} = \sum_{i=1}^M \sfrac{c_i}{\lambda} \ket{i}, \: \lambda = \sum_{i=1}^M c_i, \\
    \sel = \sum_{i=1}^M \ketbra{i}\otimes P_i,  \\
    \implies(\bra{0}\otimes I) \prep^\dag \cdot \sel \cdot \prep(\ket{0}\otimes I) = \frac{H}{\lambda},
\end{gather}
where $I$ is the identity acting on the same register as $P_i$. As such, $\prep^\dag \cdot \sel\cdot \prep$ is a \emph{block-encoding} of $H$~\cite{Low2019hamiltonian}.

In~\cite{babbush2018encoding}, the FH Hamiltonian is transformed into a LCU, with $P_i$'s being Pauli operators, using the Jordan-Wigner (JW) transformation~\cite{jw}. Then, an energy estimation algorithm is constructed by applying quantum phase estimation (QPE) to a walk operator $\mathcal{W}$, which shares the same eigenvalues as $e^{i \arccos{(H/\lambda)}}$ and $e^{-i \arccos{(H/\lambda)}}$, and can be constructed from a block-encoding and a multiply controlled $Z$ gate~\cite{PhysRevLett.121.010501,babbush2018encoding}. Further shown in~\cite{babbush2018encoding} is the computational resource estimation of the algorithm in terms of T-count and logical-qubit count. In particular, the authors estimated the total cost of the algorithm $C$ as
\begin{equation}
    C \simeq Q_{\mathcal{W}}\cdot C_{\mathcal{W}},
\end{equation}
where $Q_{\mathcal{W}}$ and $C_{\mathcal{W}}$ are the query count of $\mathcal{W}$ and the cost per $\mathcal{W}$, respectively. The costs of other parts of the algorithm are negligible and thus, not estimated in~\cite{babbush2018encoding}. We show in Supplementary Note 1 how $Q_{\mathcal{W}}$ and $C_{\mathcal{W}}$ are determined.

In~\cite{campbell2021early}, energy estimation is performed by applying QPE to the time-evolution operator $e^{-iH\tau}$, which is approximated by the second-order Trotter formula, i.e.,
\begin{equation}\label{eq:trot}
    e^{-iH\tau} \approx \left[\left ( \prod_{l=1}^{K-1} e^{-i H_l \frac{\tau}{2r}}\right) e^{-i H_K \frac{\tau}{r}}\left ( \prod_{l=K-1}^{1} e^{-i H_l \frac{\tau}{2r}} \right)\right]^r,
\end{equation}
where $r$ is the number of Trotter steps, $H = \sum_{l=1}^K H_l$ and $H_l$'s are Hermitian operators. Then, the total cost of the algorithm is
\begin{equation}
    C \simeq Q_{T}\cdot C_{T},
\end{equation}
where $Q_T$ and $C_T$ are the query count of and cost per approximated $e^{-iH\tau}$, respectively.

The FH Hamiltonian is divided into three $H_l$'s in~\cite{campbell2021early}; $H_1=H_c$, and $H_h$ is decomposed into $H_2$ and $H_3$, each consisting of hopping terms that act on non-overlapping plaquettes, which are related by a simple lattice translation, as shown in figure~\ref{fig:fig1}(a). This Trotter scheme comes with two main benefits. First, the translational symmetry between $H_2$ and $H_3$ simplifies, in practice, the evaluation of the Trotter error bounds~\cite{schubert2023trotter}. Second, every plaquette term can be diagonalized using Clifford gates and four two-site fermionic Fourier transforms~\cite{kivlichan2020improvedfault}, turning $e^{-i H_{2/3} \theta}$ into a layer of same-angle $R_z$ rotation gates. Instead of synthesizing the $R_z$ gates individually, they are more efficiently synthesized collectively using a circuit optimization technique known as \emph{Hamming-weight phasing} (HWP)~\cite{Gidney2018halvingcostof,nam2019low}. Furthermore, a chemical potential shift $n\mapsto n-1/2$ is applied to $H_c$, which leads to $3\times$ fewer Pauli operators, arising from JW transformation~\cite{jw}, in $H_c$; the resulting change in the final energy estimate can be classically corrected~\cite{campbell2021early}. Upon a Clifford transformation, $e^{-i H_c \theta}$, a product of $e^{-i ZZ\theta}$ terms acting on disjoint pairs of qubits, is reduced to a layer of same-angle $R_z$ gates, which are again effected using HWP. 

\section*{Data availability}
All data generated or analyzed during this study are included in this published article and its supplementary information files.

\section*{Code availability}
The underlying code for this study is not publicly available but may be made available to qualified researchers on reasonable request from the corresponding author.

\section*{Acknowledgements}
A. K. would like to acknowledge William Simon for his contribution on the FH simulations during the early stage of this work, Athena Caesura for optimizing the FH qubitization circuits from an earlier version of this manuscript, and Harriet Apel for spotting a few non-fatal errors, which did not affect the results, in the appendices of an earlier version of this manuscript.

\section*{Author Contributions}
B. S. wrote the code for evaluating the Trotter error bounds. A. K. designed the quantum circuits and performed the resource estimation. Both authors developed the Trotter schemes, and wrote the paper.

\section*{Competing interests}
All authors declare no financial or non-financial competing interests.

\nocite{}
\bibliographystyle{apsrev4-2}
\bibliography{ref}

\newpage

\begin{figure}
    \centering
    \includegraphics[width=\linewidth]{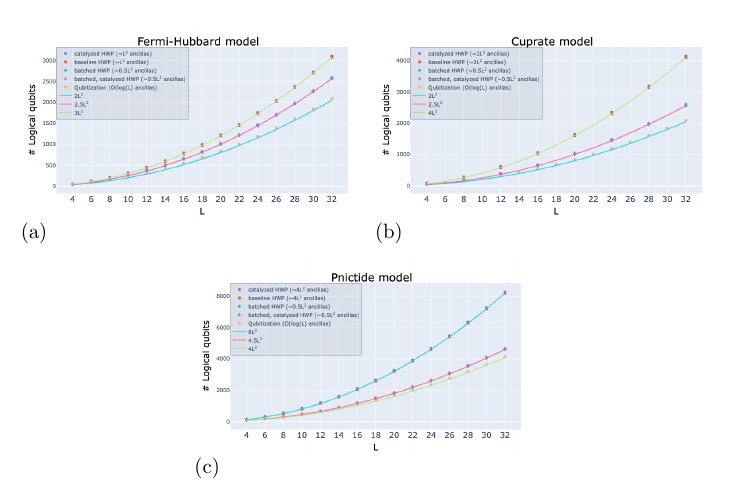}
    \caption{\textbf{Logical qubit counts for the same simulations, of which the Toffoli counts are shown in figure~\ref{fig:fig2}.} The numerical values used in these plots can be found in Supplementary Note 1 and Supplementary Note 2. We plot the solid curves as a guide for the eye.}
    \label{fig:fig3}
\end{figure}

\end{document}


\section*{Supplementary Notes for: Resource-optimized fault-tolerant simulation of the Fermi-Hubbard model and high-temperature superconductor models}

\section{Resource estimation for Qubitization algorithms}
\label{app:qubitization}

\subsection{Overview}

We present a high-level circuit construction of the qubitization algorithm, used here, from~\cite{babbush2018encoding}, as well as our error analysis and resource estimation pipeline. Given a Hamiltonian $H$, the algorithm applies a quantum phase estimation (QPE) algorithm to the walk operator $\mathcal{W}$, according to the following circuit:
\begin{equation}
\begin{adjustbox}{width = 0.85 \textwidth}
\begin{quantikz}
    \lstick{$\ket{0}$} & \gate[5]{A} & \ctrl{5}& &&&&&&\dots&&&&\gate[5]{QFT^\dag}&\meter{}\\
    \lstick{$\ket{0}$} & & &\ctrl[open]{4}&&\ctrl[open]{4}&&&&\dots&&&&&\meter{}\\
    \lstick{$\ket{0}$} & & & &&&\ctrl[open]{3}&&\ctrl[open]{3}&\dots&&&&&\meter{}\\
    \wave & & & &&&&&&&&&&&\\
    \lstick{$\ket{0}$} & & & &&&&&&\dots&\ctrl[open]{1}&&\ctrl[open]{1}&&\meter{}\\
&&\gate{\mathcal{W}}&\gate{R}&\gate{\mathcal{W}}&\gate{R}&\gate{R}&\gate{\mathcal{W}^2}&\gate{R}&\dots&\gate{R}&\gate{\mathcal{W}^{2^m-1}}&\gate{R}&&
\end{quantikz}
\end{adjustbox},
\label{eq:qb_qpe}
\end{equation}
where $A$ prepares a window state to improve the QPE success probability~\cite{greenaway2024casestudyqsvtassessment} and $QFT^\dag$ is the inverse quantum Fourier transform. The (controlled) walk operator $\mathcal{W}$ and reflection operator $R$ are realized with the circuit:
\begin{equation}\label{eq:walk}
\begin{adjustbox}{width = 0.85 \textwidth}
\begin{quantikz}
    && \ctrl{1}&\\
    \lstick{$\ket{a}$} &\qwbundle{}&\gate[2]{\mathcal{W}}&\\
    \lstick{$\ket{\psi}$} &\qwbundle{}& &
\end{quantikz}
=
\begin{quantikz}
    && \ctrl{1}&\ctrl{1}&\\
    \lstick{$\ket{a}$} &\qwbundle{}&\gate[2]{\sel}&\gate[2]{R}&\\
    \lstick{$\ket{\psi}$} &\qwbundle{}& & &
\end{quantikz}
=
\begin{quantikz}
    && \ctrl{1}&&\gate{Z}&&\\
    \lstick{$\ket{a}$} &\qwbundle{}&\gate[2]{\sel}&\gate{\prep^\dag}&\ctrl[open]{-1}&\gate{\prep}&\\
    \lstick{$\ket{\psi}$} &\qwbundle{}& & & & &
\end{quantikz}
\end{adjustbox},
\end{equation}
where given a Hamiltonian $H$ expressed as a linear combination of unitaries (LCU)~\cite{LCU,Low2019hamiltonian}:
\begin{equation}
    H = \sum_{i=1}^{M} c_i P_i, \: c_i \in \mathbbm{R}, \: c_i \geq 0
\end{equation}
with $P_i$ being unitary and self-inverse. $\sel$ and $\prep$ are used to block-encode~\cite{Low2019hamiltonian} $H$ as follows:
\begin{gather}
    \prep \ket{0} = \sum_{i=1}^M \sfrac{c_i}{\lambda} \ket{i}, \: \lambda = \sum_{i=1}^M c_i, \\
    \sel = \sum_{i=1}^M \ketbra{i}\otimes P_i  \\
    \implies(\bra{0}\otimes I) \prep^\dag \cdot \sel \cdot \prep(\ket{0}\otimes I) = \frac{H}{\lambda}.
\end{gather}
As noted in~\cite{babbush2018encoding}, since both $A$ and $QFT^\dag$ have gate complexity $\tilde{O}(m)$~\cite{nam2020approximate}, where $m$ is the number of phase qubits in the QPE, whereas $\mathcal{W}$ and $R$ have query complexity $O(2^m)$, the resource costs of $A$ and $QFT^\dag$ are completely neglected. Hence, they are ignored in the following discussion and analysis.

As in~\cite{babbush2018encoding}, we measure the simulation error through the root-mean-square error of the estimated phase $\Delta \phi$, which is related to the error budget in energy $\Delta E$ via
\begin{equation}
    \Delta E = \lambda \Delta \cos(\phi) \leq \lambda \Delta \phi.
\end{equation}
There are two sources of error: the Holevo variance~\cite{holevo2011probabilistic,PhysRevA.80.052114} of the phase estimation output $\epsilon_{QPE} = \frac{\pi}{2^{m+1}}$~\cite{babbush2018encoding}, and the circuit synthesis error of the walk operator $\epsilon_{W}$, i.e.,
\begin{equation}
    \Big(\frac{\Delta E}{\lambda}\Big)^2 \leq (\Delta \phi)^2 = \epsilon_{QPE}^2 + \epsilon_{W}^2.
\end{equation}
We split the total error budget according to
\begin{equation}\label{eq:budget}
 \epsilon_{QPE}^2  = x \Big(\frac{\Delta E}{\lambda}\Big)^2, \: \epsilon_{W}^2 = (1-x)\Big(\frac{\Delta E}{\lambda}\Big)^2,
\end{equation}
where $0<x<1$. Then, we can compute the sufficient number of phase qubits as
\begin{equation}\label{eq:FH_phase_qb}
    \frac{\pi}{2^{m+1}} = \frac{\sqrt{x} \Delta E}{\lambda} \: \implies \: m = \left\lceil \log_2\left(\frac{\pi \lambda}{2\sqrt{x}\Delta E}\right)\right\rceil < \log_2\left(\frac{\pi \lambda}{\sqrt{x}\Delta E}\right).
\end{equation}
Then, according to the circuit in~\eqref{eq:qb_qpe}, we need at most
\begin{equation}\label{eq:W_query}
    2^m < \frac{\pi \lambda}{\sqrt{x} \Delta E}
\end{equation}
queries to the walk operator to estimate the energy to within error $\Delta E$. Therefore, one can estimate the gate count of the algorithm by multiplying this query count with the gate count of the walk operator, i.e., one $\sel$ and two $\prep$, where the comparatively much smaller cost of the multiply controlled $Z$ in~\eqref{eq:walk} is neglected~\cite{babbush2018encoding}.

In what follows, we describe the low-level circuit implementation of the $\sel$ and $\prep$ oracles, and derive the resource estimates for simulating the Fermi-Hubbard model, one-band cuprate model, and two-band pnictide model. We adopt the Clifford + T + Toffoli gate set, as in~\cite{campbell2021early}, and assume that 2 T gates can be performed using 1 Toffoli gates, i.e., 2 T = 1 Toffoli, with catalyst-assisted circuits from~\cite{Gidney2019efficientmagicstate}.

\subsection{Fermi-Hubbard model}

The Fermi-Hubbard model can be expressed as
\begin{equation}\label{eq:FH_Ham}
    H = t\sum_{\langle\bi,\bj\rangle, \sigma} a^\dag_{\bi,\sigma}a_{\bj,\sigma} + u\sum_{\bi,\sigma\neq\sigma'} \left(n_{\bi,\sigma}-\frac{1}{2}\right)\left(n_{\bi,\sigma'}-\frac{1}{2}\right),
\end{equation}
where the notation $\langle\bi,\bj\rangle$ denotes nearest neighbors, and $\sigma \in \{\uparrow,\downarrow\}$ labels the spin. Upon Jordan-Wigner (JW) transformation~\cite{jw},
\begin{equation}\label{eq:FH_Hamq}
    H = \frac{t}{2}\sum_{\langle\bi,\bj\rangle,\sigma} (X_{\bi,\sigma}\vec{Z}X_{\bj,\sigma} + Y_{\bi,\sigma}\vec{Z}Y_{\bj,\sigma}) + \frac{u}{4}\sum_{\bi,\sigma \neq \sigma'} Z_{\bi,\sigma}Z_{\bi,\sigma'},
\end{equation}
where $\vec{Z}$ denotes the product of $Z$ operators that could appear in $d>1$ dimensions due to JW transformation. In the following analysis, we will assume a default JW ordering that follows a serpentine path, where fermions at the same site with opposite spins are neighbors, i.e., spin is the fastest changing variable when traversing the JW path. We consider the Fermi-Hubbard model on an $L\times L$ lattice with periodic boundary conditions. 

We first provide circuit implementations that are almost the same as those in~\cite{babbush2018encoding}, with minor differences arising from the $-1/2$ shift in the number operators. The $\prep$ oracle is defined by
\begin{align}
    &\prep \ket{0} \nonumber \\
    &\quad = \sum_{i_x,i_y=0}^{L-1}\Biggl[ \sqrt{\frac{u}{4\lambda}} \ket{1}_U \ket{i_x}\ket{i_y}\ket{0}_\sigma\ket{i_x}\ket{i_y}\ket{1}_{\sigma'} \nonumber \\
    &\quad \quad + \sfrac{t}{2\lambda}\sum_{\sigma \in \{\uparrow,\downarrow\}}\ket{0}_U\ket{i_x}\ket{i_y}\ket{\sigma}\Bigl( \ket{i_x+1}\ket{i_y}+\ket{i_x}\ket{i_y+1}+\ket{i_x-1}\ket{i_y}+\ket{i_x}\ket{i_y-1}\Bigr)\ket{\sigma} \Biggr],
\end{align}
where $i_x$ and $i_y$ are components of $\bi$ in the $x$- and $y$-directions, and 
\begin{equation}
    \lambda = 4L^2 t + \frac{u L^2}{4}.
\end{equation}
We implement this oracle using the circuit:
\begin{equation}
\centering
\begin{adjustbox}{width = 0.65 \textwidth}
\begin{quantikz}
\lstick{$\ket{0}_U$} && \gate{R_y\left(2\arccos{\sfrac{2tL^2}{\lambda}}\right)}&\ctrl[open]{6}&&&&&\ctrl[open]{4}&&\ctrl[open]{1}&\ctrl[open]{2}& \rstick{$U$}\\
\lstick{$\ket{0}_{i_x}$} &\qwbundle{\lceil \log_2 L\rceil}& \gate{USP_L}&&\ctrl{3}&&&&&&\swap{3}&&\rstick{$i_x$}\\
\lstick{$\ket{0}_{i_y}$} &\qwbundle{\lceil \log_2 L\rceil}& \gate{USP_L}&&&\ctrl{3}&&&&&&\swap{3}&\rstick{$i_y$}\\
\lstick{$\ket{0}_\sigma$} && &\gate{H}&&&\ctrl{3}&&&&&&\rstick{$\sigma$}\\
\lstick{$\ket{0}_{j_x}$} &\qwbundle{\lceil\log_2 L\rceil}& &&\targ{}&&&\swap{1}&\gate{+1\mbox{ (mod L)}}&\swap{1}&\targX{}&&\rstick{$j_x$}\\
\lstick{$\ket{0}_{j_y}$} &\qwbundle{\lceil\log_2 L\rceil}& &&&\targ{}&&\targX{}&&\targX{}&&\targX{}&\rstick{$j_y$}\\
\lstick{$\ket{0}_{\sigma'}$} && \targ{}&\targ{}&&&\targ{}&&&&&&\rstick{$\sigma'$}\\
\lstick{$\ket{0}$} && \gate{H}&&&&&\ctrl{-2}&&\ctrl{-2}&&&\\
\lstick{$\ket{0}$} && \gate{H}&&&&&&&&\ctrl{-4}&\ctrl{-3}&
\end{quantikz}
\end{adjustbox},
\label{eq:prep_FH0}
\end{equation}
The $\sel$ oracle satisfies the equation:
\begin{equation}
    \sel \ket{U,i_x,i_y,\sigma,j_x,j_y,\sigma'}\ket{\psi} = \ket{U,i_x,i_y,\sigma,j_x,j_y,\sigma'}\otimes \begin{cases}
        Z_{\bi,\sigma}Z_{\bj,\sigma'}\ket{\psi} & U \land ( \bi=\bj ) \land (\sigma = 0) \land (\sigma' = 1)\\
        X_{\bi,\sigma}\vec{Z}X_{\bj,\sigma} \ket{\psi}& \neg U \land (\bi<\bj) \land (\sigma = \sigma') \\
        Y_{\bi,\sigma}\vec{Z}Y_{\bj,\sigma} \ket{\psi}& \neg U \land (\bi>\bj) \land (\sigma = \sigma') \\
        \mbox{Undefined} & \text{otherwise,}
    \end{cases}
\end{equation}
where, when $\bi$ and $\bj$ are compared, they are treated as two-digit integers with radix $L$, i.e., $\bi = i_x + i_y L$. We implement controlled $\sel$ using the circuit:
\begin{equation}
\centering
\begin{adjustbox}{width = 0.65 \textwidth}
\begin{quantikz}
    \lstick{control} & & \ctrl{2} &&&&\ctrl{5}&&&\gate{S}& \ctrl{1}&\\
    \lstick{$U$} & & &\ctrl{1}&\ctrl{2}&\ctrl{3}&&\ctrl{3}&\ctrl{2}&\ctrl{1}&\control{}&\\
    \lstick{$i_x$} &\qwbundle{\lceil \log_2 L \rceil}& \gate{In_{i_x}}&\swap{3}& & &&&&\swap{3}& &\\
    \lstick{$i_y$} &\qwbundle{\lceil\log_2 L\rceil}& \gate{In_{i_y}}\wire[u][1]{q}&&\swap{3}&&&&\swap{3}&&&\\
    \lstick{$\sigma$} & &\gate{In_\sigma}\wire[u][1]{q}&&&\swap{3}&&\swap{3}&&&&\\
    \lstick{$j_x$} &\qwbundle{\lceil\log_2 L\rceil}& & \targX{}&&&\gate{In_{j_x}}&&&\targX{}&\gate{In_{j_x}}\wire[u][4]{q}&\\
    \lstick{$j_y$} &\qwbundle{\lceil\log_2 L\rceil}& &&\targX{}&&\gate{In_{j_y}}\wire[u][1]{q}&&\targX{}&&\gate{In_{j_y}}\wire[u][1]{q}&\\
    \lstick{$\sigma'$} & & &&&\targX{}&\gate{In_{\sigma'}}\wire[u][1]{q}&\targX{}&&&&\\
    \lstick{$\ket{\psi}$}& \qwbundle{2L^2}& \gate{\vec{Z}Y_{i_x,i_y,\sigma}}\wire[u][4]{q}&&&& \gate{\vec{Z}X_{j_x,j_y,\sigma'}}\wire[u][1]{q}&&&&\gate{-Z_{j_x, j_y,1}}\wire[u][2]{q}&
\end{quantikz}
\end{adjustbox},
\label{eq:select_FH0}
\end{equation}
where we have used the ranged operation and unary iteration operation shown in figure 7 and 9 of~\cite{babbush2018encoding}; note that the `$In$' boxes denote the address portion of a QROM gadget.

Now we show that the controlled swap gates in~\eqref{eq:select_FH0} and the $\sigma'$ qubit, as well as the gates acting on it, are superfluous. We note that this optimization is used in~\cite{baysmidt2025faulttolerant} independently from this work. First, we notice that the controlled swap gates between registers $i_x$ and $j_x$, and those between registers $i_y$ and $j_y$ are unnecessary because $\bi=\bj$ when $U=1$. The controlled swap gates between qubits $\sigma$ and $\sigma'$ can also be removed, and $\sigma$ replaces $\sigma'$ as part of the address of the middle QROM. This works because (i) when $U=0$, $\sigma = \sigma'$, and (ii) when $U=1$, $\sigma$ and $\sigma'$ are swapped. This way, one can see that the $\sigma'$ qubit is redundant and the gates acting on it in~\eqref{eq:prep_FH0} can also be removed. As a result, we obtain the optimized $\sel$ circuit:
\begin{equation}
\centering
\begin{adjustbox}{width = 0.5 \textwidth}
\begin{quantikz}
    \lstick{control} &\gate{S} & \ctrl{2} &\ctrl{4}& \ctrl{1}&\\
    \lstick{$U$} & & &&\control{}&\\
    \lstick{$i_x$} &\qwbundle{\lceil \log_2 L \rceil}& \gate{In_{i_x}} &&&\\
    \lstick{$i_y$} &\qwbundle{\lceil\log_2 L\rceil}& \gate{In_{i_y}}\wire[u][1]{q}&&&\\
    \lstick{$\sigma$} & &\gate{In_\sigma}\wire[u][1]{q}&\gate{In_\sigma}\wire[d][1]{q}&&\\
    \lstick{$j_x$} &\qwbundle{\lceil\log_2 L\rceil}& &\gate{In_{j_x}}&\gate{In_{j_x}}\wire[u][4]{q}&\\
    \lstick{$j_y$} &\qwbundle{\lceil\log_2 L\rceil}& &\gate{In_{j_y}}\wire[u][1]{q}&\gate{In_{j_y}}\wire[u][1]{q}&\\
    \lstick{$\ket{\psi}$}& \qwbundle{2L^2}& \gate{\vec{Z}Y_{i_x,i_y,\sigma}}\wire[u][4]{q}& \gate{\vec{Z}X_{j_x,j_y,\sigma'}}\wire[u][1]{q}&\gate{-Z_{j_x, j_y,1}}\wire[u][2]{q}&
\end{quantikz}
\end{adjustbox},
\label{eq:select_FH}
\end{equation}
and the optimized $\prep$ circuit:
\begin{equation}
\centering
\begin{adjustbox}{width = 0.65 \textwidth}
\begin{quantikz}
\lstick{$\ket{0}_U$} && \gate{R_y\left(2\arccos{\sfrac{2tL^2}{\lambda}}\right)}&\ctrl[open]{3}&&&&&\ctrl[open]{4}&&\ctrl[open]{1}&\ctrl[open]{2}& \rstick{$U$}\\
\lstick{$\ket{0}_{i_x}$} &\qwbundle{\lceil \log_2 L\rceil}& \gate{USP_L}&&\ctrl{3}&&&&&&\swap{3}&&\rstick{$i_x$}\\
\lstick{$\ket{0}_{i_y}$} &\qwbundle{\lceil \log_2 L\rceil}& \gate{USP_L}&&&\ctrl{3}&&&&&&\swap{3}&\rstick{$i_y$}\\
\lstick{$\ket{0}_\sigma$} && &\gate{H}&&&&&&&&&\rstick{$\sigma$}\\
\lstick{$\ket{0}_{j_x}$} &\qwbundle{\lceil\log_2 L\rceil}& &&\targ{}&&&\swap{1}&\gate{+1\mbox{ (mod L)}}&\swap{1}&\targX{}&&\rstick{$j_x$}\\
\lstick{$\ket{0}_{j_y}$} &\qwbundle{\lceil\log_2 L\rceil}& &&&\targ{}&&\targX{}&&\targX{}&&\targX{}&\rstick{$j_y$}\\
\lstick{$\ket{0}$} && \gate{H}&&&&&\ctrl{-2}&&\ctrl{-2}&&&\\
\lstick{$\ket{0}$} && \gate{H}&&&&&&&&\ctrl{-4}&\ctrl{-3}&
\end{quantikz}
\end{adjustbox}.
\label{eq:prep_FH}
\end{equation}

We detail below the non-Clifford count and ancilla count of various circuit components in $\sel$ and $\prep$.
\begin{itemize}
    \item A singly controlled QROM of $N$ items costs $N-1$ Toffoli gates and $\lceil \log_2(N)\rceil$ ancilla qubits~\cite{babbush2018encoding}.
    \item Each arbitrary-angle single-qubit rotation gate about the $x/y/z$ axis can be synthesized using the repeat-until-success (RUS) method from~\cite{Kliuchnikov2023shorterquantum} with a mean T count of $0.53 \log_2(1/\delta) + 4.86$ and 1 ancilla qubit, where $\delta$ is the error budget of the rotation gate.
    \item Each $n$-controlled swap gate, i.e., $C^nSwap$ is synthesized by turning the middle CNOT gate of the standard decomposition of a swap gate into 3 CNOTs into an $n$-controlled NOT, i.e., $C^nX$ gate. 
    \item Each $n$-controlled Hadamard gate, i.e., $C^nH = S^\dag H T^\dag (C^nX) T H S$, where the single qubit rotation gates act on the target qubit of $C^nX$.
    \item Using the method from~\cite{PhysRevA.87.022328}, adding $n$ control qubits to a controlled unitary costs $n$ Toffoli gates and $n$ clean ancilla qubits in $\ket{0}$. So, a $C^nX$ gate costs $n-1$ Toffoli gates and $n-2$ ancilla qubits.
    \item Uniform state preparation (USP): We aim to prepare $\sum_{l=0}^{2^k m-1} c_l \ket{l}$, where $|c_l|^2 = \frac{1}{2^k m}$. If $m=1$, only Hadamard gates are required and no ancilla qubits are needed. For $m>1$, the amplitude-amplification method from~\cite{babbush2018encoding} or~\cite{PRXQuantum.2.030305} costs $2\lceil\log_2(m)\rceil-2$ Toffoli gates, 2 $R_z$ gates, and $\lceil\log_2(m)\rceil$ ancilla qubits for $m>1$. Converting the $R_z$ count to the expected T count using the RUS circuits from~\cite{Kliuchnikov2023shorterquantum} and then to Toffoli count according to 1 Toffoli = 2 T~\cite{Gidney2019efficientmagicstate}, this method is expected to cost $2\lceil\log_2(m)\rceil + 2(0.53\log_2(1/\epsilon)+4.86)-2$ Toffoli gates, where $\epsilon$ is the error budget per $R_z$. 
    \item Addition/Subtraction with classical numbers and between addends of different lengths: We consider an $n$-bit adder from~\cite{Gidney2018halvingcostof}, and modify its building-block according to
    \begin{equation}
    \begin{adjustbox}{width = 0.95 \textwidth}
    \begin{quantikz}[transparent]
    \lstick{$c_k$} &\ctrl{2}& &\ctrl{3}&&&\ctrl{3}&&\ctrl{1}&&\rstick{$c_k$}\\
    \lstick{$i_k$} &\targ{}&\ctrl{1}&&&&&\ctrl{2}&\targ{}&\ctrl{1}&\rstick{$i_k$}\\
    \lstick{$t_k$} &\targ{}&\ctrl{1}&&&&&\ctrl{1}&&\targ{}&\rstick{$(i+t)_k$}\\
     &\setwiretype{n}&\wire[u][1]{q}&\targ{}\setwiretype{q}&\push{c_{k+1}} \setwiretype{n}& \dots\push{c_{k+1}}&\targ{}&\wire[u][1]{q}\setwiretype{q}
    \end{quantikz}
    $\mapsto$
    \begin{quantikz}[transparent]
    \lstick{$c_k$} &\ctrl{1}&\gate{X^{i_k}}&\ctrl{1}&\gate{X^{i_k}}&\ctrl{2}&&&\ctrl{2}&\gate{X^{i_k}}&\ctrl{2}&\gate{X^{i_k}}&\ctrl{1}&\rstick{$c_k$}\\
    \lstick{$t_k$} &\targ{}&&\ctrl{1}&&&&&&&\ctrl{1}&&\targ{}&\rstick{$(i+t)_k$}\\
     &\setwiretype{n}&&\wire[u][1]{q}&\setwiretype{q}&\targ{}&\push{c_{k+1}} \setwiretype{n}& \dots\push{c_{k+1}}&\targ{}&\setwiretype{q}&\wire[u][1]{q}
    \end{quantikz}
    \end{adjustbox},
    \end{equation}
    when $i_k$ is a classically known bit, and $X^{i_k}$ is a NOT gate when $i_k=1$ and an identity otherwise. This modification works because the first CNOT between $c_k$ and $i_k$ copies $c_k (\oplus 1)$ into $i_k$ when $i_k=0\:(1)$, and hence, the subsequent Toffoli gate can be controlled by $c_k (\oplus 1)$ if $i_k=0\:(1)$ instead of $i_k$. There are two corollaries. First, when adding a classical state, i.e., a computational basis state, to a quantum state, the classical state does not need to be stored in qubits. Second, when adding two quantum states of different lengths, the shorter one does not need to be padded by 0's.
    \item Controlled addition with classical numbers: Instead of using a controlled adder, which costs $2n - 1$ Toffoli gates~\cite{Gidney2018halvingcostof}, we initialize the classical number in a quantum register using CNOTs before adding it into an quantum addend using an uncontrolled adder, before undoing the CNOTs. As such, a controlled incrementer costs $n-2$ Toffoli gates and $n$ ancilla qubits, where 1 ancilla qubit stores the classical addend 1 and $n$ is the length of the quantum addend.
    \end{itemize}

The non-Clifford gate counts per $\prep$ are 1 $R_y$ gate,  2 T gates, and $5\lceil \log_2(L)\rceil - 1$ Toffoli gates, as well as the non-Clifford gates required by the two $USP_L$'s. The non-Clifford gate count per $\sel$ is $5L^2 - 2$ Toffoli gates. Since the ancilla qubit count of a $\sel$ is larger than that of a $\prep$, the ancilla qubit count of a walk operator, i.e., size of $\ket{a}$ in \eqref{eq:walk}, is given by that of a $\sel$, which is $6 \lceil\log_2(L)\rceil + 4$. Therefore, combining with~\eqref{eq:FH_phase_qb}, the total qubit count is 
\begin{equation}
    \Big\lceil\log_2\left(\frac{\pi \lambda L^6}{2\sqrt{x}\Delta E}\right)\Big\rceil+2L^2 + 3.
\end{equation}
We multiply the gate count per walk operator, i.e., one $\sel$ and two $\prep$, by the query count in~\eqref{eq:W_query} to obtain the total respective numbers of $R_{y/z}$, T, and Toffoli gates:
\begin{gather}
    N_R = 
    \begin{cases}
        \frac{2 \pi \lambda}{\sqrt{x} \Delta E}, \mbox{ if }L\mbox{ is a binary power.}\\
        \frac{6 \pi \lambda}{\sqrt{x} \Delta E}, \mbox{ otherwise.}
    \end{cases} \\ 
    N_T = \frac{4 \pi \lambda}{\sqrt{x} \Delta E}, \\
    N_{Tof} = \begin{cases}
        \frac{\pi \lambda}{\sqrt{x} \Delta E}\left(5L^2 + 10\lceil \log_2(L)\rceil - 4 \right), \mbox{ If }L\mbox{ is a binary power.}\\
        \frac{\pi \lambda}{\sqrt{x} \Delta E}\left(5L^2 + 10\lceil \log_2(L)\rceil + 4 \lceil \log_2(m)\rceil -4 \right), \mbox{ otherwise,}    
        \end{cases} 
\end{gather}
where $m$ is the smallest integer such that $L = 2^k m$. Then, the $R_{y/z}$ gates are synthesized with Clifford + T gates using RUS circuits~\cite{Kliuchnikov2023shorterquantum}, which implies
\begin{equation}
    N_T = 
    \begin{cases}
        \frac{\pi \lambda}{\sqrt{x} \Delta E} \left( 4+2\cdot(0.53\log_2\Big( \frac{2\pi\lambda^2}{\sqrt{x(1-x)}(\Delta E)^2}\Big) + 4.86) \right), \mbox{ if }L\mbox{ is a binary power.}\\
        \frac{\pi \lambda}{\sqrt{x} \Delta E} \left( 4+6\cdot(0.53\log_2\Big( \frac{6\pi\lambda^2}{\sqrt{x(1-x)}(\Delta E)^2}\Big) + 4.86) \right), \mbox{ otherwise.}
    \end{cases}
\end{equation}
where we used~\eqref{eq:budget}, i.e., $\epsilon_W = \frac{\sqrt{1-x}\Delta E}{\lambda}$. Assuming that 1 Toffoli = 2 T~\cite{Gidney2019efficientmagicstate}, we numerically minimize the total Toffoli count, i.e., $N_{Tof} + N_T/2$, with respect to $x$, and find $x\approx 0.99$ to be near optimal, i.e., yielding the lowest Toffoli count. The total Toffoli and qubit counts for $u=8$ and $t=1$ are summarized in Supplementary Table~\ref{tb:FH_QB}.

\begin{table}[!ht]
\begin{tabular}{l|l|l}
L  & $\#$ Toffoli & $\#$ qubits \\ \hline
4  & $4.33e5$     & 58          \\ \hline
6  & $9.75e5$     & 102         \\ \hline
8  & $1.36e6$     & 160         \\ \hline
10 & $2.19e6$     & 234         \\ \hline
12 & $3.03e6$     & 324         \\ \hline
14 & $3.99e6$     & 429         \\ \hline
16 & $4.96e6$     & 550         \\ \hline
18 & $6.38e6$     & 687         \\ \hline
20 & $7.81e6$     & 840         \\ \hline
22 & $9.36e6$     & 1009        \\ \hline
24 & $1.11e7$     & 1194        \\ \hline
26 & $1.29e7$     & 1395        \\ \hline
28 & $1.50e7$     & 1611        \\ \hline
30 & $1.71e7$     & 1844        \\ \hline
32 & $1.93e7$     & 2092       
\end{tabular}
\caption{Resource estimates of qubitization simulations of the $L\times L$ Fermi-Hubbard model with $u=8$, $t=1$, and $\Delta E = 0.51\% L^2$.}
\label{tb:FH_QB}
\end{table}

\subsection{Single-orbital Cuprate model}

The cuprate model considered here extends the Fermi-Hubbard model by including second- and third-nearest-neighbor hopping terms, leading to the following Hamiltonian:
\begin{equation}\label{eq:Cu_Ham}
    H = t\sum_{\langle\bi,\bj\rangle, \sigma} a^\dag_{\bi,\sigma}a_{\bj,\sigma} + t'\sum_{\langle\langle\bi,\bj\rangle\rangle, \sigma} a^\dag_{\bi,\sigma}a_{\bj,\sigma} + t''\sum_{\langle\langle\langle\bi,\bj\rangle\rangle\rangle, \sigma} a^\dag_{\bi,\sigma}a_{\bj,\sigma} + u\sum_{\bi,\sigma\neq\sigma'} \left(n_{\bi,\sigma}-\frac{1}{2}\right)\left(n_{\bi,\sigma'}-\frac{1}{2}\right),
\end{equation}
where $t'$ and $t''$ are the hopping strength between second- and third-nearest neighbors, respectively. Post-JW transformation, the Hamiltonian becomes
\begin{gather}\label{eq:cu_Hamq}
    H = \frac{t}{2}\sum_{\langle\bi,\bj\rangle,\sigma} (X_{\bi,\sigma}\vec{Z}X_{\bj,\sigma} + Y_{\bi,\sigma}\vec{Z}Y_{\bj,\sigma}) +\frac{t'}{2}\sum_{\langle\langle\bi,\bj\rangle\rangle,\sigma} (X_{\bi,\sigma}\vec{Z}X_{\bj,\sigma} + Y_{\bi,\sigma}\vec{Z}Y_{\bj,\sigma}) \nonumber  \\ +\frac{t''}{2}\sum_{\langle\langle\langle\bi,\bj\rangle\rangle\rangle,\sigma} (X_{\bi,\sigma}\vec{Z}X_{\bj,\sigma} + Y_{\bi,\sigma}\vec{Z}Y_{\bj,\sigma}) + \frac{u}{4}\sum_{\bi,\sigma \neq \sigma'} Z_{\bi,\sigma}Z_{\bi,\sigma'},
\end{gather}
where $\vec{Z}$ denotes the product of $Z$ operators that could appear in $d>1$ dimensions due to JW transformation

The $\prep$ oracle performs the transformation:
\begin{align}\label{eq:prep_cu_def}
    &\prep \ket{0} \nonumber \\
    &\quad = \sum_{i_x,i_y=0}^{L-1}\Biggl[ \sqrt{\frac{u}{4\lambda}} \ket{1}_U \ket{0}_V\ket{0}_W\ket{i_x}\ket{i_y}\ket{0}_{\sigma}\ket{i_x}\ket{i_y}\ket{1}_{\sigma'} \nonumber \\
    &\quad \quad + \sfrac{t}{2\lambda}\sum_{\sigma \in \{\uparrow,\downarrow\}}\ket{0}_U\ket{0}_V\ket{0}_W\ket{i_x}\ket{i_y}\ket{\sigma}\Bigl( \ket{i_x+1}\ket{i_y}+\ket{i_x}\ket{i_y+1}+\ket{i_x-1}\ket{i_y}+\ket{i_x}\ket{i_y-1}\Bigr)\ket{\sigma} \nonumber \\
    &\quad \quad+ \sfrac{t'}{2\lambda}\sum_{\sigma \in \{\uparrow,\downarrow\}}\ket{0}_U\ket{1}_V\ket{0}_W\ket{i_x}\ket{i_y}\ket{\sigma}\Bigl( \ket{i_x+1}\ket{i_y+1}+\ket{i_x+1}\ket{i_y-1}+\ket{i_x-1}\ket{i_y+1}+\ket{i_x-1}\ket{i_y-1}\Bigr)\ket{\sigma} \nonumber \\
    &\quad \quad + \sfrac{t''}{2\lambda}\sum_{\sigma \in \{\uparrow,\downarrow\}}\ket{0}_U\ket{0}_V\ket{1}_W\ket{i_x}\ket{i_y}\ket{\sigma}\Bigl( \ket{i_x+2}\ket{i_y}+\ket{i_x}\ket{i_y+2}+\ket{i_x-2}\ket{i_y}+\ket{i_x}\ket{i_y-2}\Bigr)\ket{\sigma} \Biggr],
\end{align}
where
\begin{equation}
    \lambda = 4L^2(t+t'+t'') + \frac{uL^2}{4}.
\end{equation}
Its circuit implementation is:
\begin{equation}\label{eq:prep_cu}
\centering
\begin{adjustbox}{width = \textwidth}
\begin{quantikz}[transparent]
    \lstick{$\ket{0}_U$} & & \gate{R_y\left(\theta_1\right)}&\ctrl[open]{1}&&&\ctrl[open]{5}&&&\ctrl[open]{1}& & &&\ctrl[open]{1}&& &&\ctrl[open]{3}&\ctrl[open]{4}&\ctrl[open]{6}&\rstick{$U$}\\
    \lstick{$\ket{0}_V$} & & & \gate{R_y\left(\theta_2 \right)}&\ctrl{1}&\targ{}&&&&\ctrl[open]{6}&&&&\ctrl[open]{6}&&\ctrl{7}&&&&&\rstick{$V$}\\
    \lstick{$\ket{0}_W$} & & && \gate{R_y\left( \theta_3 \right)}&\ctrl{-1}&&&&&\ctrl{4}&&\ctrl{4}&&&&&&&&\rstick{$W$}\\
    \lstick{$\ket{0}_{i_x}$} &\qwbundle{\log L} &\gate{USP_{L}}&&&&&\ctrl{4}&&&&&&&&&&\swap{4}&&&\rstick{$i_x$}\\
    \lstick{$\ket{0}_{i_y}$} &\qwbundle{\log L} &\gate{USP_{L}}&&&&&&\ctrl{4}&&&&&&&&&&\swap{4}&&\rstick{$i_y$}\\
    \lstick{$\ket{0}_\sigma$} & &&&&&\gate{H}&&&&&&&&&&&&&&\rstick{$\sigma$}\\
    \lstick{$\ket{0}$} &\qwbundle{2}&&&&&\gate{\oplus 1}\wire[u][1]{q}&&&&\gate{\oplus 1}&\gate[2]{+ \mod L}&\gate{\oplus 1}&&&&&&&\gate{\oplus 1}&\rstick{$\ket{0}$}\\
    \lstick{$\ket{0}_{j_x}$} &\qwbundle{\log L}&&&&&&\targ{}&&\swap{1}&&&&\swap{1}&&\linethrough&&\targX{}&&&\rstick{$j_x$}\\
    \lstick{$\ket{0}_{j_y}$} &\qwbundle{\log L}&&&&&&&\targ{}&\targX{}&&&&\targX{}&\gate{Neg}&\gate{+1 \mod L}&\gate{Neg}&&\targX{}&&\rstick{$j_y$}\\ 
    \lstick{$\ket{0}$}& & \gate{H}&&&&&&&\ctrl{-1}&&&&\ctrl{-1}&&&&&&& \\
    \lstick{$\ket{0}$}& & \gate{H}&&&&&&&&&&&&\ctrl{-2}&&\ctrl{-2}&&&& \\
    \lstick{$\ket{0}$}& & \gate{H}&&&&&&&&&&&&&&&\ctrl{-6}&\ctrl{-5}&& 
\end{quantikz}
\end{adjustbox},
\end{equation}
where $\theta_1 = 2\arcsin{\sqrt{\frac{L^2 u}{2\lambda}}}$, $\theta_2 = 2\arccos{\sqrt{(1+\frac{t}{t'}+\frac{t}{t''})^{-1}}}$, and $\theta_3=2\arccos{\sqrt{(1+\frac{t'}{t''})^{-1}}}$.
We provide some clarifications and explanation about the components in this circuit that are absent in~\eqref{eq:prep_FH}.
\begin{itemize}
    \item The first four gates acting on registers $U,V,W$ initialize them in the correct bit patterns and amplitudes to be distributed by the other gates, as per~\eqref{eq:prep_cu_def}.
    \item The controlled $R_y$ gates cost two uncontrolled $R_y$ gates and two CNOT gates~\cite{10.1145/1120725.1120847}.
    \item The first two $C^3 Swap$ gates are compiled using the identity
    \begin{equation}\label{eq:ladder}
    \centering
    \begin{adjustbox}{width = 0.5 \textwidth}
    \begin{quantikz}[transparent]
    \lstick{$x_1$}&\ctrl{3}&&\ctrl{3}& \\
    \lstick{$x_2$}&\ctrl{2}&&\ctrl{2}& \\
    \lstick{$x_3$}&\ctrl{1}&&\ctrl{1}& \\
    \lstick{$x_4$}&\gate[2]{V}&&\gate[2]{V}& \\
    \lstick{$x_5$}&&\gate[2]{U}&& \\
    \lstick{$x_6$}&&&& \\
    \end{quantikz}
    =
    \begin{quantikz}[transparent]
    \lstick{$x_1$}&\ctrl{2}&&&&&&\ctrl{2}&\\
    \lstick{$x_2$}&\ctrl{1}&&&&&&\ctrl{1}&\\
    \setwiretype{n}& &\ctrl{2}\setwiretype{q}&&&&\ctrl{2}&\wire[u]{q}\\
    \lstick{$x_3$}& &\ctrl{1}& &&&\ctrl{1}&&\\
    \setwiretype{n}& & &\ctrl{1}\setwiretype{q}& &\ctrl{1}&\\
    \lstick{$x_4$}& & &\gate[2]{V} & &\gate[2]{V}&&&\\
    \lstick{$x_5$}& &&&\gate[2]{U}&&&&\\
    \lstick{$x_6$}& &&&&&&&\\
    \end{quantikz}
    \end{adjustbox},
    \end{equation}
    where $V$ is taken to be a swap gate.
    \item The first controlled $\oplus 1$ operation is a CNOT that initializes an ancilla register to $\ket{01} = \ket{1}$ to be added to register $j_x$ for generating the second and third lines in~\eqref{eq:prep_cu_def}. The second controlled $\oplus 1$ operation consists of two CNOTs that turn $\ket{01}$ to $\ket{10} = \ket{2}$ to be added to register $j_x$ for generating the fourth line in~\eqref{eq:prep_cu_def}. The third and fourth controlled $\oplus 1$ operations uncompute the second and first ones, respectively.
    \item The pair of controlled $Neg$ operations are CNOTs that turn the adder conjugated by them into a subtractor~\cite{cuccaro2004new}.
\end{itemize}
The $\sel$ implementation is the same as that in~\eqref{eq:select_FH}.

The non-Clifford gate counts per $\prep$ are 5 $R_y$ gates, 2 T gates, and $6\lceil \log_2(L) \rceil + 2$ Toffoli gates, as well as the non-Clifford gates required by the $USP_L$'s. The ancilla qubit count is the same as that of a $\sel$, so the total qubit count is
\begin{equation}
    \Big\lceil\log_2\left(\frac{\pi \lambda L^6}{2\sqrt{x}\Delta E}\right)\Big\rceil+2L^2 + 3.
\end{equation}
The T and Toffoli counts are
\begin{gather}
    N_T = 
    \begin{cases}
        \frac{\pi \lambda}{\sqrt{x} \Delta E}\left( 4+ 10\cdot(0.53\log_2\Big( \frac{10\pi\lambda^2}{\sqrt{x(1-x)}(\Delta E)^2}\Big) + 4.86)\right), & \mbox{ if }L\mbox{ is a binary power.}\\
        \frac{\pi \lambda}{\sqrt{x} \Delta E}\left( 4+ 14\cdot(0.53\log_2\Big( \frac{14\pi\lambda^2}{\sqrt{x(1-x)}(\Delta E)^2}\Big) + 4.86)\right),&\mbox{ otherwise.}
    \end{cases}\\
    N_{Tof} = \begin{cases}
        \frac{\pi \lambda}{\sqrt{x} \Delta E} (5L^2+ 12 \lceil \log_2(L)\rceil+ 2),& \mbox{ if }L\mbox{ is a binary power.}\\
        \frac{\pi \lambda}{\sqrt{x} \Delta E} (5L^2+ 12 \lceil \log_2(L)\rceil + 4\lceil \log_2(m)\rceil + 2), & \mbox{ otherwise.}
    \end{cases}
\end{gather}
As in the Fermi-Hubbard simulations, we numerically minimize the total Toffoli count, i.e., $N_{Tof} + N_T/2$, with respect to $x$, and find $x\approx 0.99$ to be near optimal. We set $u=8$, $t=1$, $t'=0.3$, and $t''=0.2$, which are in range of the parameter settings used in condensed matter literature~\cite{delannoy2009low,dalla2012unified,Hirayama2018ab,moree2022ab,lebert2023paramagnon}. We further choose $\Delta E = 0.51\% L^2$, assuming that this model is as hard to simulate classically as the Fermi-Hubbard model. Though, we believe it is likely harder and so, a larger $\Delta E$ will likely suffice for a quantum advantage; we leave for future work to determine this $\Delta E$ value. The total Toffoli and qubit counts are summarized in Supplementary Table~\ref{tb:cu_QB}.

\begin{table}[!ht]
\begin{tabular}{l|l|l}
L  & $\#$ Toffoli & $\#$ qubits \\ \hline
4  & $1.04e6$     & 59          \\ \hline
6  & $1.78e6$     & 102         \\ \hline
8  & $2.29e6$     & 161         \\ \hline
10 & $3.41e6$     & 235         \\ \hline
12 & $4.53e6$     & 324         \\ \hline
14 & $5.81e6$     & 430         \\ \hline
16 & $7.10e6$     & 551         \\ \hline
18 & $9.01e6$     & 688         \\ \hline
20 & $1.09e7$     & 841         \\ \hline
22 & $1.30e7$     & 1010        \\ \hline
24 & $1.53e7$     & 1194        \\ \hline
26 & $1.78e7$     & 1395        \\ \hline
28 & $2.05e7$     & 1612        \\ \hline
30 & $2.33e7$     & 1844        \\ \hline
32 & $2.62e7$     & 2093       
\end{tabular}
\caption{Resource estimates of qubitization simulations of the $L\times L$ cuprate model with $u=8$, $t=1$, $t'=0.3$, and $t''=0.2$, and $\Delta E = 0.51\% L^2$.}
\label{tb:cu_QB}
\end{table}

\subsection{Two-orbital Pnictide model}

We consider the two-orbital Pnictide model from~\cite{Raghu2008minimal} defined by
\begin{gather}
    H = t_1 \left[\sum_{\langle\bi,\bj\rangle_{\be{y}},\sigma} a^{\dag}_{\bi,\sigma,x} a_{\bi,\sigma,x} + \sum_{\langle\bi,\bj\rangle_{\be{x}},\sigma} a^{\dag}_{\bi,\sigma,y} a_{\bi,\sigma,y} \right] + t_2 \left[\sum_{\langle\bi,\bj\rangle_{\be{x}},\sigma} a^{\dag}_{\bi,\sigma,x} a_{\bi,\sigma,x} + \sum_{\langle\bi,\bj\rangle_{\be{y}},\sigma} a^{\dag}_{\bi,\sigma,y} a_{\bi,\sigma,y} \right] \nonumber \\
    + t_3 \sum_{\langle\langle\bi,\bj\rangle\rangle,\sigma,d} a^{\dag}_{\bi,\sigma,d} a_{\bi,\sigma,d} + t_4\left[ \sum_{\langle\langle\bi,\bj\rangle\rangle_{\be{x}+\be{y}},\sigma,d\neq d'} a^{\dag}_{\bi,\sigma,d} a_{\bi,\sigma,d'}- \sum_{\langle\langle\bi,\bj\rangle\rangle_{\be{x}-\be{y}},\sigma,d\neq d'} a^{\dag}_{\bi,\sigma,d} a_{\bi,\sigma,d'}\right] \nonumber \\
    + u\sum_{\bi, \sigma\neq\sigma',d} \left(n_{\bi,\sigma,d}-\frac{1}{2}\right)\left(n_{\bi,\sigma',d}-\frac{1}{2}\right)+ v\sum_{\bi, \sigma,\sigma',d\neq d'} \left(n_{\bi,\sigma,d}-\frac{1}{2}\right)\left(n_{\bi,\sigma',d'}-\frac{1}{2}\right),
\end{gather}
where $\langle\bi,\bj\rangle_{\be{x}/\be{y}}$ denotes nearest neighbors in the $\be{x}/\be{y}$ direction, $\langle\langle\bi,\bj\rangle\rangle_{\be{x}\pm\be{y}}$ denotes the next-nearest neighbors in the $\be{x}\pm\be{y}$ direction, $d\in \{x,y\}$ labels the fermionic orbitals, and $u$ and $v$ are the strengths of the intra- and inter-orbital on-site Coulomb interaction. We do not consider other weaker types of on-site interaction, such as Hund's rule coupling and pair hopping that are over an order of magnitude weaker than Coulomb interactions~\cite{Raghu2008minimal,moreo2009properties}. After JW transformation, the Hamiltonian becomes
\begin{gather}
    H =  \sum_{P\in\{X,Y\}}\Bigg\{\frac{t_1}{2}\left[\sum_{\langle\bi,\bj\rangle_{\be{y}},\sigma} P_{\bi,\sigma,x}\vec{Z}P_{\bi,\sigma,x} \sum_{\langle\bi,\bj\rangle_{\be{x}},\sigma} P_{\bi,\sigma,y} \vec{Z} P_{\bi,\sigma,y} \right] + \frac{t_2}{2} \left[\sum_{\langle\bi,\bj\rangle_{\be{x}},\sigma} P_{\bi,\sigma,x}\vec{Z} P_{\bi,\sigma,x} + \sum_{\langle\bi,\bj\rangle_{\be{y}},\sigma} P_{\bi,\sigma,y} \vec{Z} P_{\bi,\sigma,y} \right] \nonumber \\
    + \frac{t_3}{2} \sum_{\langle\langle\bi,\bj\rangle\rangle,\sigma,d} P_{\bi,\sigma,d}\vec{Z}P_{\bi,\sigma,d} + \frac{t_4}{2}\left[ \sum_{\langle\langle\bi,\bj\rangle\rangle_{\be{x}+\be{y}},\sigma,d\neq d'} P_{\bi,\sigma,d} \vec{Z}P_{\bi,\sigma,d'}- \sum_{\langle\langle\bi,\bj\rangle\rangle_{\be{x}-\be{y}},\sigma,d\neq d'} P_{\bi,\sigma,d} \vec{Z}P_{\bi,\sigma,d'}\right] \Bigg\}\nonumber \\
    + \frac{u}{4}\sum_{\bi, \sigma\neq\sigma',d} Z_{\bi,\sigma,d}Z_{\bi,\sigma',d}+ \frac{v}{4}\sum_{\bi, \sigma,\sigma',d\neq d'} Z_{\bi,\sigma,d}Z_{\bi,\sigma',d'}.
\end{gather}
Note that while fermions with opposite spins at the same site are adjacent to each other along the JW path, fermions with different $d$'s, i.e., orbitals, are not. For a given $d$-value, as stated above, we choose the JW path to be a serpentine path, where spin is the fastest-changing variable; we join the end of the path for $d=x$ (represented by $d=0$ on a quantum computer) with the beginning of the path for $d=y$ (represented by $d=1$ on a quantum computer) to obtain the full JW path. So, $d$ is the slowest changing variable.

The $\prep$ operation performs the transformation:
{\allowdisplaybreaks
\begin{align}\label{eq:prep_pn_def}
    &\prep \ket{0} \nonumber \\
    &\quad = \sum_{i_x,i_y=0}^{L-1} \Bigg\{ \sfrac{v}{4\lambda} \sum_{\sigma,\sigma'\in\{\uparrow,\downarrow\}} \ket{1}_U\ket{1}_V\ket{0}_W\ket{0}_Q\ket{i_x}\ket{i_y}\ket{\sigma}\ket{0}_{d}\ket{i_x}\ket{i_y}\ket{\sigma'}\ket{1}_{d'} \nonumber \\
    &\quad + \sfrac{u}{4\lambda} \sum_{d\in\{x,y\}} \ket{1}_U\ket{0}_V\ket{0}_W\ket{0}_Q \ket{i_x}\ket{i_y}\ket{0}_\sigma \ket{d}\ket{i_x}\ket{i_y}\ket{1}_{\sigma'}\ket{d} \nonumber \\
    &\quad + \sfrac{t_1}{2\lambda} \sum_{\sigma\in\{\uparrow,\downarrow\}} \ket{0}_U\ket{0}_V\ket{0}_W\ket{0}_Q\ket{i_x}\ket{i_y}( \ket{0}_d\ket{\sigma}\ket{i_x}\ket{i_y+1}\ket{0}_{d'}\ket{\sigma} + \ket{1}_d\ket{\sigma}\ket{i_x+1}\ket{i_y}\ket{1}_{d'}\ket{\sigma} \nonumber \\
    &\quad + \ket{0}_d\ket{\sigma}\ket{i_x}\ket{i_y-1}\ket{0}_{d'}\ket{\sigma}+ \ket{1}_d\ket{\sigma}\ket{i_x-1}\ket{i_y}\ket{1}_{d'}\ket{\sigma}) \nonumber\\
    &\quad + \sfrac{t_2}{2\lambda} \sum_{\sigma\in\{\uparrow,\downarrow\}} \ket{0}_U\ket{1}_V\ket{0}_W\ket{0}_Q\ket{i_x}\ket{i_y}(\ket{0}_d\ket{\sigma}\ket{i_x+1}\ket{i_y}\ket{0}_{d'}\ket{\sigma}+\ket{1}_d\ket{\sigma}\ket{i_x}\ket{i_y+1}\ket{1}_{d'}\ket{\sigma}\nonumber \\
    &\quad +\ket{0}_d\ket{\sigma}\ket{i_x-1}\ket{i_y}\ket{0}_{d'}\ket{\sigma}+\ket{1}_d\ket{\sigma}\ket{i_x}\ket{i_y-1}\ket{1}_{d'}\ket{\sigma})\nonumber\\
    &\quad + \sfrac{t_3}{2\lambda} \sum_{\sigma\in\{\uparrow,\downarrow\}} \sum_{d\in \{x,y\}}\ket{0}_U\ket{0}_V\ket{1}_W\ket{0}_Q \ket{i_x}\ket{i_y} \ket{d}\ket{\sigma}(\ket{i_x+1}\ket{i_y+1} +\ket{i_x+1}\ket{i_y-1} \nonumber \\
    &\quad + \ket{i_x-1}\ket{i_y-1} + \ket{i_x-1}\ket{i_y+1})\ket{d}\ket{\sigma}\nonumber \\ 
    &\quad + \sfrac{t_4}{2\lambda} \sum_{\sigma\in\{\uparrow,\downarrow\}} \Big[ \ket{0}_U\ket{1}_V\ket{1}_W\ket{0}_Q \ket{i_x}\ket{i_y}(\ket{0}_d\ket{\sigma}\ket{i_x+1}\ket{i_y+1}\ket{1}_{d'}\ket{\sigma}+ \ket{1}_d\ket{\sigma}\ket{i_x+1}\ket{i_y+1}\ket{0}_{d'}\ket{\sigma}\nonumber \\
    &\quad + \ket{0}_d\ket{\sigma}\ket{i_x-1}\ket{i_y-1}\ket{1}_{d'}\ket{\sigma}+\ket{1}_d\ket{\sigma}\ket{i_x-1}\ket{i_y-1}\ket{0}_{d'}\ket{\sigma}) \nonumber \\
    &\quad + \ket{0}_U\ket{0}_V\ket{1}_W\ket{1}_Q \ket{i_x}\ket{i_y}(\ket{0}_d\ket{\sigma}\ket{i_x+1}\ket{i_y-1}\ket{1}_{d'}\ket{\sigma}+\ket{1}_d\ket{\sigma}\ket{i_x+1}\ket{i_y-1}\ket{0}_{d'}\ket{\sigma}+\nonumber \\
    &\quad \ket{0}_d\ket{\sigma}\ket{i_x-1}\ket{i_y+1}\ket{1}_{d'}\ket{\sigma}+\ket{1}_d\ket{\sigma}\ket{i_x-1}\ket{i_y+1}\ket{0}_{d'}\ket{\sigma})\Big] \Bigg\}, 
\end{align}}
where
\begin{equation}
    \lambda = L^2\left[4(t_1 + t_2) + 8(t_3+t_4) + \frac{u}{2} + v\right].
\end{equation}
Its circuit implementation is:
\begin{equation}
    \centering
    \begin{adjustbox}{width = \textwidth}
    \begin{quantikz}[transparent]
    \lstick{$\ket{0}_U$} & \gate[4]{Load}&&\ctrl{1}&\ctrl{1}&\ctrl[open]{6}&&&&&\ctrl{1}&&&\ctrl[open]{1}&\ctrl[open]{1}&&&&\ctrl[open]{9}&&&&\ctrl[open]{1}&\ctrl[open]{1}&&&&&&\rstick{$U$}\\
    \lstick{$\ket{0}_V$} & &&\ctrl{5}&\ctrl{9}&&&&&&\ctrl{5}&\ctrl{1}&&\ctrl{1}&\ctrl[open]{1}&&\ctrl{1}&&&&\ctrl{1}&&\ctrl[open]{1}&\ctrl{1}&&&&&&\rstick{$V$}\\
    \lstick{$\ket{0}_W$} & &&&&&&&&&&\ctrl{9}&\ctrl{1}&\ctrl[open]{5}&\ctrl[open]{5}&&\ctrl{7}&\ctrl{1}&&\ctrl{1}&\ctrl{7}&&\ctrl[open]{5}&\ctrl[open]{5}&\ctrl{1}&\ctrl{6}&\ctrl{1}&\ctrl{2}&\ctrl{3}&\rstick{$W$}\\
    \lstick{$\ket{0}_Q$} & &&&&&&&&&&&\ctrl{8}&&&&&\ctrl{6}&&\ctrl{6}&&&&&\ctrl{5}&&\ctrl{5}&&&\rstick{$Q$}\\
    \lstick{$\ket{0}_{i_x}$} &\qwbundle{\log L}&\gate{USP_L}&&&&\ctrl{4}&&&&&&&&&&&&&&&&&&&&&\swap{4}&&\rstick{$i_x$}\\
    \lstick{$\ket{0}_{i_y}$} &\qwbundle{\log L}&\gate{USP_L}&&&&&\ctrl{4}&&&&&&&&&&&&&&&&&&&&&\swap{4}&\rstick{$i_y$}\\
    \lstick{$\ket{0}_{\sigma}$} &&&\gate{H}\wire[d]{q}&&\gate{H}\wire[d][4]{q}&&&\ctrl{4}&&\ctrl{4}&&&&&&&&&&&&&&&&&&&\rstick{$\sigma$}\\
    \lstick{$\ket{0}_{d}$} &\gate{H}&&\gate{H}\wire[d][3]{q}&&&&&&\ctrl{4}&&&&\ctrl[open]{1}&\ctrl{1}&&&&&&&&\ctrl{1}&\ctrl[open]{1}&&&&&&\rstick{$d$}\\
    \lstick{$\ket{0}_{j_x}$} &\qwbundle{\log L}&\gate{USP_L}&&&&\targ{}&&&&&&&\swap{1}&\swap{1}&&&&&&&&\swap{1}&\swap{1}&\gate{Neg}&\gate{+1 \mod L}&\gate{Neg}&\targX{}&&\rstick{$j_x$}\\
    \lstick{$\ket{0}_{j_y}$} &\qwbundle{\log L}&\gate{USP_L}&&&&&\targ{}&&&&&&\targX{}&\targX{}&\gate{Neg}&\gate{Neg}&\gate{Neg}&\gate{+1 \mod L}&\gate{Neg}&\gate{Neg}&\gate{Neg}&\targX{}&\targX{}&&&&&\targX{}&\rstick{$j_y$}\\
    \lstick{$\ket{0}_{\sigma'}$} &\targ{}&&\targ{}&\gate{H}\wire[d][1]{q}&\targ{}&&&\targ{}&&\targ{}&&&&&&&&&&&&&&&&&&&\rstick{$\sigma'$}\\
    \lstick{$\ket{0}_{d'}$} & &&&\targ{}&&&&&\targ{}&&\targ{}&\targ{}&&&&&&&&&&&&&&&&&\rstick{$d'$}\\
    \lstick{$\ket{0}$} &\gate{H}&&&&&&&&&&&&&&\ctrl{-3}&\ctrl{-3}&\ctrl{-3}&&\ctrl{-3}&\ctrl{-3}&\ctrl{-3}&&&&&&&&\\
    \lstick{$\ket{0}$} &\gate{H}&&&&&&&&&&&&&&&&&&&&&&&&&&\ctrl{-5}&\ctrl{-4}&
    \end{quantikz}
    \end{adjustbox},
\end{equation}
where
\begin{equation}
    \centering
    \begin{adjustbox}{width = 0.75 \textwidth}
    \begin{quantikz}[transparent]
    \lstick{$\ket{0}_U$}&\gate[4]{Load}& \\
    \lstick{$\ket{0}_V$}&&\\
    \lstick{$\ket{0}_W$}&& \\
    \lstick{$\ket{0}_Q$}&&
    \end{quantikz}
    =
    \begin{quantikz}[transparent]
    \lstick{$\ket{0}_U$}&\gate{R_y(\theta_1)}& \ctrl{1}&\ctrl[open]{1}&\ctrl[open]{1}&&&&&&\\
    \lstick{$\ket{0}_V$}&&\gate{R_y(\theta_2)}&\gate{R_y(\theta_3)}&\ctrl{1}&&&&\targ{}&\ctrl{1}&\\
    \lstick{$\ket{0}_W$}&&&&\gate{R_y(\theta_4)}&\ctrl{1}&\targ{}&\gate{H}&\ctrl{-1}&\swap{1}&\\
    \lstick{$\ket{0}_Q$}&&&&&\gate{R_y(\theta_5)}&\ctrl{-1}&\ctrl{-1}&&\targX{}&
    \end{quantikz}
    \end{adjustbox},
\end{equation}
where $\theta_1 = 2\arcsin{\sfrac{L^2(u+v)}{2\lambda}}$, $\theta_2 = 2\arcsin{\sqrt{(1+\frac{v}{u})^{-1}}}$, $\theta_3 = 2\arccos{\sqrt{(1+\frac{t_1}{t_2}+\frac{t_1}{2t_3}+\frac{t_1}{2t_4})^{-1}}}$, $\theta_4=2\arccos$ ${\sqrt{(1+\frac{t_2}{2t_3}+\frac{t_2}{2t_4})^{-1}}}$, and $\theta_5 = 2\arccos{\sqrt{(1+\frac{t_3}{t_4})^{-1}}}$.
Note that we apply~\eqref{eq:ladder} to compile the multi-controlled swap gates and negation operations that conjugate the controlled incrementers. For other multi-controlled gates, we store the logical conjunctions in clean ancilla qubits, whenever they can be used in future operations; these logical conjunctions are uncomputed when they are no longer needed for future operations, using only Clifford gates and measurements~\cite{Gidney2018halvingcostof}. For example, $U \land V$ is computed using a Toffoli into a clean ancilla qubit; it is then XOR'ed into the target qubits in the decompositions of the Hadamard and NOT gates controlled by $\ket{U}$ and $\ket{V}$, and its logical AND with $\ket{\sigma}$ is computed into $\ket{\sigma'}$ for the NOT gate controlled by $\ket{U}$, $\ket{V}$, and $\ket{\sigma}$. Moreover, we compile the multi-controlled rotation gate using the identity~\cite{Wang2021resourceoptimized}
\begin{equation}
    \centering
    \begin{adjustbox}{width = 0.35 \textwidth}
    \begin{quantikz}[transparent]
    \lstick{$x_1$}&\ctrl{1}&\\
    \lstick{$x_2$}&\ctrl{1}&\\
    \lstick{$x_3$}&\gate{R_z(\theta)}&
    \end{quantikz}
    =
    \begin{quantikz}[transparent]
    \lstick{$x_1$}&\ctrl{1}&&&&\ctrl{1}&\\
    \lstick{$x_2$}&\ctrl{2}&&&&\ctrl{2}&\\
    \lstick{$x_3$}& & \ctrl{1}&\gate{R_z(\theta/2)}&\ctrl{1}&&\\
    \setwiretype{n}&&\targ{}\setwiretype{q}&\gate{R_z(-\theta/2)}&\targ{}&&\setwiretype{n}
    \end{quantikz}
    \end{adjustbox}.
\end{equation}

The $\sel$ operation is defined by
\begin{align}
    &\sel\ket{U,V,W,Q,i_x,i_y,\sigma,d,j_x,j_y,\sigma',d'}\ket{\psi} \nonumber \\
    &= \ket{U,V,W,Q,i_x,i_y,\sigma,d,j_x,j_y,\sigma',d'}\otimes \begin{cases}
        Z_{\bi,\sigma,d} Z_{\bj,\sigma',d'} \ket{\psi}& U\land (\bi=\bj)\\
        X_{\bi,\sigma,d} \vec{Z} X_{\bj,\sigma',d'} \ket{\psi} & \neg U\land\neg (W\land Q)\land (M(d,\bi)<M(d',\bj)) \land(\sigma=\sigma') \\
        Y_{\bi,\sigma,d} \vec{Z} Y_{\bj,\sigma',d'} \ket{\psi} & \neg U\land\neg (W\land Q)\land (M(d,\bi)>M(d',\bj)) \land(\sigma=\sigma')\\
        -X_{\bi,\sigma,d} \vec{Z} X_{\bj,\sigma',d'} \ket{\psi} & W\land Q \land (M(d,\bi)<M(d',\bj)) \land(\sigma=\sigma') \\
        -Y_{\bi,\sigma,d} \vec{Z} Y_{\bj,\sigma',d'} \ket{\psi} & W\land Q \land (M(d,\bi)>M(d',\bj)) \land(\sigma=\sigma')\\
        \mbox{Undefined} & \text{otherwise,}
    \end{cases}
\end{align}
where $M(d,\bi) = i_x + i_y L + d L^2$. Note that $\vec{Z}$ in the first case reduces to an identity when $d = d'$.
The controlled $\sel$ circuit is:
\begin{equation}
    \centering
    \begin{adjustbox}{width = 0.5 \textwidth}
    \begin{quantikz}[transparent]
    \lstick{control}&&\ctrl{1}&\ctrl{1}&\gate{S}&\ctrl{3}&\ctrl{1}&\ctrl{1}&\\
    \lstick{$U$}&&\ctrl[open]{4}&\ctrl[open]{8}&\ctrl[open]{-1}&&\ctrl{1}&\ctrl{1}&\\
    \lstick{$V$}&&&&&&\ctrl{5}&\ctrl[open]{3}&\\
    \lstick{$W$}&&&&&\ctrl{1}&&&\\
    \lstick{$Q$}&&&&&\control{}&&&\\
    \lstick{$i_x$}&\qwbundle{\log L}&\gate{In_{i_x}}\wire[d]{q}&&&&&\gate{In_{i_x}}\wire[d]{q}&\\
    \lstick{$i_y$}&\qwbundle{\log L}&\gate{In_{i_y}}\wire[d]{q}&&&&&\gate{In_{i_y}}\wire[d][2]{q}&\\
    \lstick{$\sigma$}&&\gate{In_{\sigma}}\wire[d]{q}&&&&\gate{In_{\sigma}}\wire[d][2]{q}&&\\
    \lstick{$d$}&&\gate{In_{d}}&&&&&\gate{In_{d}}&\\
    \lstick{$j_x$}&\qwbundle{\log L}&&\gate{In_{j_x}}\wire[d]{q}&&&\gate{In_{j_x}}\wire[d]{q}&&\\
    \lstick{$j_y$}&\qwbundle{\log L}&&\gate{In_{j_y}}\wire[d]{q}&&&\gate{In_{j_y}}\wire[d]{q}&&\\
    \lstick{$\sigma'$}&&&\gate{In_{\sigma'}}\wire[d]{q}&&&\gate{In_{\sigma'}}\wire[d]{q}&&\\
    \lstick{$d'$}&&&\gate{In_{d'}}\wire[d]{q}&&&\wire[d]{q}&&\\
    \lstick{$\ket{\psi}$}&\qwbundle{4L^2}&\gate{\vec{Z}Y_{i_x,i_y,\sigma,d}}\wire[u][5]{q}&\gate{\vec{Z}X_{j_x,j_y,\sigma',d'}}&&&\gate{Z_{\bj,\sigma,0} Z_{\bj,\sigma',1}}&\gate{Z_{\bi,0,d} Z_{\bi,1,d}}\wire[u][5]{q}&
    \end{quantikz}
    \end{adjustbox}.
\end{equation}

The non-Clifford gate counts per $\prep$ are 9 $R_y$ gates, 10 T gates, and $12+8 \lceil\log_2 L \rceil$ Toffoli gates, as well as the non-Clifford gates required by the $USP_L$'s. The non-Clifford gate counts per $\sel$ are $14L^2+3$ Toffoli gates and 2 T gates. The total qubit count is given by
\begin{equation}
    \left\lceil \log_2\left(\frac{\pi \lambda L^6}{2\sqrt{x}\Delta E}\right)\right\rceil +4L^2 + 10.
\end{equation}
The T and Toffoli counts are
\begin{gather}
    N_T = 
    \begin{cases}
        \frac{\pi \lambda}{\sqrt{x} \Delta E}\left( 22+ 18\cdot(0.53\log_2\Big( \frac{18\pi\lambda^2}{\sqrt{x(1-x)}(\Delta E)^2}\Big) + 4.86)\right),& \mbox{ if }L\mbox{ is a binary power.}\\
        \frac{\pi \lambda}{\sqrt{x} \Delta E}\left( 22+ 22\cdot(0.53\log_2\Big( \frac{22\pi\lambda^2}{\sqrt{x(1-x)}(\Delta E)^2}\Big) + 4.86)\right),&\mbox{ otherwise.}
    \end{cases}\\
    N_{Tof} = \begin{cases}
        \frac{\pi \lambda}{\sqrt{x} \Delta E} (14L^2+ 12 \lceil \log_2(L)\rceil+ 27),& \mbox{ if }L\mbox{ is a binary power.}\\
        \frac{\pi \lambda}{\sqrt{x} \Delta E} (14L^2+ 12 \lceil \log_2(L)\rceil + 4\lceil \log_2(m)\rceil + 27),&\mbox{ otherwise.}
    \end{cases}
\end{gather}

Once again, we numerically minimize the total Toffoli count, i.e., $N_{Tof} + N_T/2$, with respect to $x$, and find $x\approx 0.99$ to be near optimal. We set $t_1=1$, $t_2=1.3$, $t_3=0.85$, and $t_4=0.85$, following~\cite{Raghu2008minimal}. We further choose $u=8$ and $\Delta E = 0.51\% L^2$, assuming that $u/t_1 = 8$ is the point where this model is classically as hard to simulate as the Fermi-Hubbard model with $u/t=8$. We believe this assumption is very lenient as the system size of this model is double that of the Fermi-Hubbard model. The total Toffoli and qubit counts are summarized in Supplementary Table~\ref{tb:pn_QB}.

\begin{table}[!ht]
\begin{tabular}{l|l|l}
L  & $\#$ Toffoli & $\#$ qubits \\ \hline
4  & $1.08e7$     & 100          \\ \hline
6  & $1.81e7$     & 183         \\ \hline
8  & $2.56e7$     & 298         \\ \hline
10 & $3.76e7$     & 444         \\ \hline
12 & $5.11e7$     & 621         \\ \hline
14 & $6.67e7$     & 831         \\ \hline
16 & $8.39e7$     & 1072         \\ \hline
18 & $1.05e8$     & 1345        \\ \hline
20 & $1.29e8$     & 1650        \\ \hline
22 & $1.54e8$     & 1987        \\ \hline
24 & $1.82e8$     & 2355        \\ \hline
26 & $2.12e8$     & 2756        \\ \hline
28 & $2.45e8$     & 3189        \\ \hline
30 & $2.79e8$     & 3653        \\ \hline
32 & $3.16e8$     & 4150       
\end{tabular}
\caption{Resource estimates of qubitization simulations of the $L\times L$ pnictide model with $u=8$, $t_1=1$, $t_2=1.3$, $t_3=0.85$, and $t_4=0.85$, and $\Delta E = 0.51\% L^2$.}
\label{tb:pn_QB}
\end{table}

\section{Resource estimation for Trotter algorithms}
\label{app:trotter}

\subsection{Overview}

We present a high-level overview of the Trotter-based energy-estimation algorithm, used here, from~\cite{kivlichan2020improvedfault}, along with our error analysis and resource estimation pipeline. Specifically, given a Hamiltonian $H$, this algorithm applies the quantum phase estimation algorithm from~\cite{higgins2007entanglement} to query the directionally-controlled time evolution, i.e., $\ket{c}\ket{\psi} \mapsto e^{-i (-1)^c H\tau}\ket{c}\ket{\psi}$, instead of the controlled time evolution; the number of queries $N_{q}$ to reach a root-mean-squared error of $\Delta \phi$ is~\cite{kivlichan2020improvedfault}
\begin{equation}
    N_q \approx \frac{0.76 \pi}{\Delta \phi},
\end{equation}
where $\Delta \phi$ translates to an energy error $\Delta E_{PE}$ via
\begin{equation}
    \Delta \phi = \Delta E_{PE} \cdot \tau.
\end{equation}
The time-evolution is approximated using the symmetric, second-order Trotter formula~\cite{suzuki1991general}
\begin{equation}\label{eq:trot2}
    e^{-iH\tau} \approx U(\tau,r) \equiv \Bigg[\left ( \prod_{l=1}^{K-1} e^{-i H_l \frac{\tau}{2r}}\right) e^{-i H_K \frac{\tau}{r}}\left ( \prod_{l=K-1}^{1} e^{-i H_l \frac{\tau}{2r}} \right)\Bigg]^r,
\end{equation}
where the Hamiltonian is decomposed into an ordered sum: $H = \sum_{i=1}^K H_i$ and $H_i$'s are Hermitian, and $r$ is the number of Trotter steps. Note that $r$ is set to 1 in~\cite{kivlichan2020improvedfault,campbell2021early}. The directionally-controlled time evolution for the Hamiltonians considered in this work only incurs an additional Clifford overhead compared to the uncontrolled time-evolution~\cite{kivlichan2020improvedfault}. 

In addition to $\Delta E_{PE}$, there are two additional sources of errors: (i) Trotter approximation error $\Delta E_T$ and (ii) rotation synthesis error $\Delta E_R$; note that we have expressed the errors in the same unit as $\Delta E_{PE}$. The total error budget is given by
\begin{gather}
    \Delta E = \Delta E_{PE} + \Delta E_T + \Delta E_R, \mbox{ where} \\
    \Delta E_{PE} = y \Delta E,\: \Delta E_T = (1-s)(1-y) \Delta E, \: \Delta E_R = s(1-y)\Delta E, \mbox{ with }0<s,y<1. \label{eq:trotbudget}
\end{gather}
$\Delta E_T$ is upper-bounded by~\cite{kivlichan2020improvedfault}
\begin{equation}\label{eq:trot_err}
    \Delta E_T \tau \leq || e^{-iH\tau} - U(\tau,r)|| \equiv \delta, \mbox{ if } \delta \ll 1,
\end{equation}
where $||\cdot ||$ is the spectral norm. We bound the Trotter error $\delta$ by computing the commutator bounds~\cite{childs2021theory}:
\begin{equation}\label{eq:trot_err2}
    \delta \leq \frac{\tau^3}{12r^2}\sum_{i=1}^K \Bigg|\Bigg| \Bigg[ \Bigg[H_i, \sum_{j=i+1}^K H_j\Bigg], \sum_{k=i+1}^K H_k\Bigg] \Bigg|\Bigg| + \frac{\tau^3}{24r^2}\sum_{i=1}^K \Bigg|\Bigg| \Bigg[ \Bigg[H_i,\sum_{j=i+1}^K H_j\Bigg], H_i\Bigg] \Bigg|\Bigg| \equiv \frac{\tau^3 W}{r^2}.
\end{equation}
We compute $r$ using~\eqref{eq:trotbudget}-\eqref{eq:trot_err2}:
\begin{equation}\label{eq:rW}
    \Delta E_T \tau = \frac{\tau^3 W}{r^2} \: \implies \: r = \Bigg\lceil \sfrac{W\tau^2}{(1-s)(1-y)\Delta E} \Bigg \rceil.
\end{equation}
Finally, the gate count of the algorithm is estimated by multiplying $N_q$ with the gate count per $U(\tau,r)$. In practice, the Toffoli count is optimized over the parameters, $s,y$, and $\tau$, while keeping the total error $\leq \Delta E$ and $\delta \ll 1$. 

\subsection{Fermi-Hubbard model}

\noindent \textbf{Trotter scheme and error:} We adopt the Trotter scheme, i.e., the ordered decomposition of $H$, dubbed PLAQ, and its error analysis from~\cite{campbell2021early}, which we present below. We assume an $L\times L$ lattice with an even $L$ and periodic boundary conditions, and denote it by $\Lambda = (\mathbbm{Z}_L)^2 = \{ (0,0),(0,1),(1,0),(1,1),...,(L-1,L-1)\}$. We also define a sub-lattice $\Lambda' = ((2\mathbbm{Z})_L)^2 = \{(0,0),(0,2),(2,0),(2,2),...,(L-2,L-2) \}$. Furthermore, we let $p_1=(0,0)$, $p_2=(1,0)$, $p_3=(1,1)$ and $p_4=(0,1)$. We denote a hopping operator as
\begin{equation}
    h_{i,j,\sigma} \equiv a^\dag_{i,\sigma}a_{j,\sigma}+a^\dag_{j,\sigma}a_{i,\sigma}.
\end{equation}
Then, the nearest-neighbor (nn) hopping term of the Hamiltonian is decomposed into two terms:
\begin{equation}
    H_{nn,1} = t\sum_{\bi \in \Lambda'}\sum_{\sigma}\sum_{k=1}^4 h_{\bi+p_k,\bi+p_{k+1},\sigma}, \\
    H_{nn,2} = t\sum_{\bi \in \Lambda'}\sum_{\sigma}\sum_{k=1}^4 h_{\bi+p_k-p_3,\bi+p_{k+1}-p_3,\sigma},
\end{equation}
where $H_{nn,1}$ and $H_{nn,2}$ are equivalent up to a lattice translation by $(1,1)$. The on-site Coulomb term, given by
\begin{equation}\label{eq:FH_on}
    H_{c} = u\sum_{\bi \in \Lambda} \left(n_{\bi,\uparrow}-\frac{1}{2}\right)\left(n_{\bi,\downarrow}-\frac{1}{2}\right),
\end{equation}
is implemented as one term. The order in which these three terms are evolved in the Trotter formula is defined by the ordered list:
\begin{equation}
    \{H_i\}_{i=1}^3 = \{ H_c,H_{nn,1},H_{nn,2}\}.
\end{equation}
As reported in~\cite{campbell2021early}, the Trotter error is bounded from above by:
\begin{equation}
    \frac{\delta r^2}{\tau^3} \leq  \frac{|ut^2L^2|}{6}(\sqrt{5}+8) + \frac{|u^2 t|}{24} ||H_{nn,1}+H_{nn,2}|| + \frac{3}{24}||[ [H_{nn,1},H_{nn,2}],H_{nn,1}]||,
\end{equation}
where the last two terms can be efficiently computed classically because $H_{nn,1}$ and $H_{nn,2}$ are free-fermionic; we display their computed values, which we take from~\cite{campbell2021early}, in Supplementary Table~\ref{tb:fh_bounds}.

\begin{table}[!ht]
\begin{tabular}{l|l|l|l|l|l|l|l|l|l|l|l|l|l|l|l}
$L$                                       & 4  & 6   & 8   & 10  & 12  & 14  & 16  & 18  & 20  & 22  & 24  & 26   & 28   & 30   & 32   \\ \hline
$||H_{nn,1}+H_{nn,2}||/t$                 & 24 & 56  & 100 & 160 & 230 & 320 & 410 & 520 & 650 & 780 & 930 & 1100 & 1300 & 1500 & 1700 \\ \hline
$||[ [H_{nn,1},H_{nn,2}],H_{nn,1}]||/t^3$ & 0  & 110 & 190 & 300 & 440 & 630 & 810 & 1000 & 1300 & 1600 & 1800 & 2200  & 2500  & 2900  & 3300  \\ 
\end{tabular}
\caption{Numerically computed values of $||H_{nn,1}+H_{nn,2}||/|t|$ and $||[ [H_{nn,1},H_{nn,2}],H_{nn,1}]||/|t^3|$ at various values of $L$ for an $L\times L$, periodic Fermi-Hubbard model, taken from~\cite{campbell2021early}.}
\label{tb:fh_bounds}
\end{table}

\noindent \textbf{Circuit implementation:}
Our circuit implementation of $e^{-iH_i \tau}$ is the same as that in~\cite{campbell2021early} up to rotation synthesis, which we discuss below. We perform the same JW transformation as in the qubitization case.

\underline{On-site term $e^{-i H_c \theta}$}: After JW transformation, $e^{-i H_c \theta}$ is implemented using $L^2$ $e^{-i ZZ\theta}$ gates applied to pairs of qubits that represent fermions on the same site. Each $e^{-i ZZ\theta}$ is equivalent to one $R_z(2\theta)$ gate conjugated by a pair of CNOT gates~\cite{nielsen2010quantum}. 

In~\cite{campbell2021early}, the $L^2$ same-angle $R_z(\theta)$ gates can be applied using Hamming-weight phasing (HWP)~\cite{Gidney2018halvingcostof}, a circuit optimization technique widely used in quantum simulations~\cite{nam2019low,kan2022lattice,kan2022simulating,kivlichan2020improvedfault,campbell2021early}. In HWP, the Hamming-weight of the to-be-rotated $L^2$-qubit state is first computed into an ancilla register of size $\lfloor \log_2(L^2) \rfloor + 1$. Then, $\lfloor \log_2(L^2) \rfloor + 1$ $R_z$ rotations $\bigotimes_{i=0}^{\lfloor \log_2(L^2) \rfloor} R_z(2^i \theta)$ are applied to the ancilla register, before the Hamming weight is uncomputed. The Hamming weight of a binary register can be computed using $L^2 - w(L^2)$ half- and full-adders~\cite{muller1975bounds}, where $w(L^2)$ computes the number of 1s in the binary representation of $L^2$. A half- or full-adder can be implemented, up to Clifford gates, with a Toffoli gate, and can be uncomputed via a measurement-feed-forward Clifford circuit~\cite{Gidney2018halvingcostof}. See Supplementary Figure~\ref{fig:Hamming_weight} for the half- and full-adder circuits, and an example of an 8-bit Hamming-weight computation circuit. This implies that the computation and un-computation of the Hamming-weight function cost $L^2-w(L^2)$ Toffoli gates.

We optimize HWP by implementing $\bigotimes_{i=0}^{\lfloor \log_2(L^2) \rfloor} R_z(2^i \theta)$ using a generalized phase-gradient operation, shown in Supplementary Figure~\ref{fig:phase_gadget}, instead of direct synthesis; we dub this gate-optimized HWP \emph{catalyzed HWP}. While baseline HWP requires $L^2-w(L^2)$, where $w(L^2)$ computes the number of 1's in the binary representation of $L^2$, Toffoli gates and $\lfloor \log_2(L^2)\rfloor + 1$ $R_z$ rotations, catalyzed HWP requires $L^2+\lfloor \log_2(L^2)\rfloor -w(L^2)+1$ Toffoli gates and only one $R_z$, as well as a reusable catalyst state $\bigotimes_{i=0}^{\lfloor \log_2(L^2) \rfloor} R_z(2^i \theta)\ket{+...+}$ whose one-time synthesis costs are amortized over the course of the simulation.

\underline{NN hopping term $e^{-i H_{nn,i}\theta}$}: The implementation of $e^{-i H_{nn,1}\theta}$ and $e^{-i H_{nn,2}\theta}$ are very similar. Both $H_{nn,1}$ and $H_{nn,2}$ consist of hopping operators that act on the sites surrounding non-overlapping plaquettes. Inheriting the labeling choice made in~\cite{campbell2021early}, we choose to label pairwise interacting fermions around a plaquette such that their fermionic labels, defined by a fermion-to-qubit transformation, differ by $1\mod 4$. However, no explicit circuit construction was provided in~\cite{campbell2021early}.
We accomplish this by first performing a JW transformation with a serpentine ordering shown in Supplementary Figure~\ref{fig:jw}. Then, we apply an appropriate fermionic swap network, which we will describe shortly, to achieve the desired fermionic ordering.

Before getting to the fermionic swap network, we must first define a fermionic swap:
\begin{equation}\label{eq:f_swap}
    f^{swap}_{i,j} a^{\dag}_j (f^{swap}_{i,j})^\dag = a^\dag_i, \: f^{swap}_{i,j} a_j (f^{swap}_{i,j})^\dag = a_i.
\end{equation}
Assuming a JW transformation, a nearest-neighbor (nn) fermionic swap $f^{swap}_{i,i+1}$ can be implemented as a swap followed by a CZ to correct the phase. An operation that shifts the $i$th fermion to the $(i+k)$th site can be composed using a series of nn fermionic swap, i.e.,
\begin{equation}\label{eq:f_shift}
    f_{i \rightarrow i+k} \equiv f^{swap}_{i+k-1,i+k}...f^{swap}_{i+1,i+2}f^{swap}_{i,i+1}, \: f_{i \rightarrow i+k} a_i (f_{i \rightarrow i+k})^\dag = a_{i+k}.
\end{equation}
Then, we can swap the $i$th and $(i+k)$th fermions by first shifting the $i$th fermion to the $(i+k)$th site and then, the $(i+k-1)$th fermion to the $i$th site, i.e.,
\begin{equation}\label{eq:nonlocal_fswap}
    f^{swap}_{i,i+k} = f_{i+k-1 \rightarrow i}f_{i \rightarrow i+k}
\end{equation}
which costs $2k-1$ nn fermionic swaps.

We begin with the swap network for $H_{nn,1}$. For a given spin and for each row of plaquettes, we label the plaquettes $1$ through $p = \frac{L}{2}$, where $L$ is the number of sites along a dimension, from left to right. We then pair up the plaquettes $(1,2), (3,4), ..., (p-1,p)$ if $p$ is even; if $p$ is odd, then the last pair will be $(p-2, p-1)$ and the $p$th plaquette is left unpaired. Finally, for each plaquette pair, we swap the bottom left fermion of the left plaquette with the top right fermion of the right plaquette, and the bottom right fermion of the left plaquette with the top left fermion of the right plaquette. This costs $O(L^3)$ nn fermionic swaps because there are $O(L^2)$ plaquette pairs and $O(L)$ nn fermionic swaps are performed on each pair. This guarantees that the desired ordering of fermions around every plaquette. As for $H_{nn,2}$, we first map it to $H_{nn,1}$ by shifting the sites in the left-most column to the right by $L-1$ sites, and those in the bottom-most row upwards by $L-1$ sites; then, we apply the swap network for $H_{nn,1}$. Note that shifting a site upwards by a site requires $O(L^2)$ nn fermionic swaps. So, localizing the plaquettes in $H_{nn,2}$ to the desired fermionic ordering also costs $O(L^3)$ nn fermionic swaps. 

\begin{figure}
    \centering
    \includegraphics[width = 0.3\textwidth]{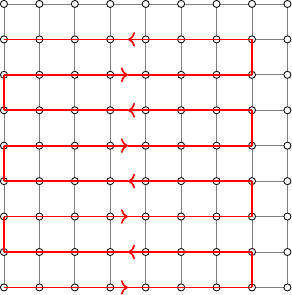}
    \caption{A serpentine path for Jordan-Wigner transformation. The top-most and right-most sites are identified with the bottom-most and left-most sites due to periodic boundary conditions.}
    \label{fig:jw}
\end{figure}

Then, each of $e^{-i H_{nn,1}\theta}$ and $e^{-i H_{nn,2}\theta}$ will be transformed into $L^2/2$, where a factor of 2 accounts for the two spin sector, plaquette evolutions of the form $e^{i \theta K_i}$ with~\cite{campbell2021early}
\begin{gather}
    K_i = 2 (b^\dag_i b_i - c^\dag_i c_i) \\
    b_i = (a_{i+1} + a_{i+2} + a_{i+3} + a_{i+4})/2 \\ 
    c_i = (a_{i+1} - a_{i+2} + a_{i+3} - a_{i+4})/2.
\end{gather}
These unitaries act on disjoint plaquettes, indicated by the index $i$. Hereafter, we assume WLOG that $i=0$. Let $F_{j,k}$ be a two-site fermionic Fourier transform:
\begin{equation}
    F_{j,k} a_j F_{j,k}^\dag = \frac{(a_j + a_k)}{\sqrt{2}}, \: F_{j,k} a_k F_{j,k}^\dag = \frac{(a_j - a_k)}{\sqrt{2}}.
\end{equation}
Then, we define
\begin{equation}\label{eq:V_op}
    V = f^{swap}_{2,3} F_{1,2} F_{3,4} f^{swap}_{1,2} F_{2,3},
\end{equation}
Using the fact that
\begin{equation}
    Va_2 V^\dag = b, \: Va_3 V^\dag = c,
\end{equation}
we can realize a plaquette evolution as
\begin{align}
    e^{i \theta K} &= V e^{i 2\theta a^\dag_2 a_2} e^{i 2\theta a^\dag_3 a_3} V^\dag \\
    &= f^{swap}_{2,3} F_{1,2} F_{3,4} f^{swap}_{1,2} F_{2,3}e^{i 2\theta a^\dag_2 a_2} e^{i 2\theta a^\dag_3 a_3} F_{2,3} f^{swap}_{1,2}F_{3,4}F_{1,2} f^{swap}_{2,3}. \nonumber
\end{align}
Assuming a JW transformation, $F_{2,3}e^{i 2\theta a^\dag_2 a_2} e^{i 2\theta a^\dag_3 a_3} F_{2,3} = e^{i\theta X_2 X_3}e^{i\theta Y_2 Y_3}$~\cite{campbell2021early}, which can be mapped to a single layer of two same-angle $R_z(2 \theta)$ gates acting on qubits 2 and 3 using Clifford gates~\cite{Wang2021resourceoptimized}. Each $F$ operation requires two T gates. Thus, each $e^{i\theta K}$ costs eight T gates and two $R_z$ gates. As a result, each of $e^{-i \theta H_{nn,1}}$ and $e^{-i \theta H_{nn,2}}$ can be implemented using $4L^2$ T gates and a layer of $L^2$ same-angle $R_z$ gates, which we realize using catalyzed HWP once again.

\noindent\textbf{Resource estimation:} The total gate count is given by the gate count per approximated evolution $U(\tau,r)$ multiplied by the number of queries $N_q$ to it. To construct $U(\tau,r)$, $e^{-i\theta H_c}$ is applied $r+1$ times, $e^{-i\theta H_{nn,1}}$ is applied $2r$ times, and $e^{-i\theta H_{nn,2}}$ is applied $r$ times, where adjacent evolutions of the same term are merged; note that each term is implemented using catalyzed HWP, the costs of which are delineated in the main text (see ``Space-time trade-off in Hamming-weight phasing"), up to Clifford gates and two-site fermionic Fourier transforms. Each $e^{-i\theta H_c}$ costs 1 $R_z$ gate and $L^2 + \lfloor \log_2(L^2)\rfloor- w(L^2) +1$ Toffoli gates, and each $e^{-i\theta H_{nn,1}}$ or $e^{-i\theta H_{nn,2}}$ costs 1 $R_z$ gate, $4L^2$ T gates, and $L^2 + \lfloor \log_2(L^2)\rfloor- w(L^2) +1$ Toffoli gates, not including the synthesis of the catalyst states. The required catalyst states are $\bigotimes_{i=0}^{\lfloor \log_2(L^2) \rfloor+1} R_z(2^i \frac{u \tau}{2r}) \ket{+...+}$ for $e^{-i\frac{u \tau}{4r} H_c}$ and $e^{-i\frac{u \tau}{2r} H_c}$ that arises from merging two adjacent $e^{-i\frac{u \tau}{4r} H_c}$'s, and $\bigotimes_{i=0}^{\lfloor \log_2(L^2) \rfloor + 1} R_z(2^i \frac{t \tau}{r}) \ket{+...+}$ for $e^{-i\frac{t \tau}{2r} H_{nn,1}}$, $e^{-i\frac{t \tau}{2r} H_{nn,2}}$, and $e^{-i\frac{t \tau}{r} H_{nn,2}}$ that arises from merging two adjacent $e^{-i\frac{t \tau}{2r} H_{nn,2}}$'s; these states cost $2\lfloor \log_2(L^2) \rfloor + 4$ $R_z$ gates.

The Toffoli count is
\begin{equation}\label{eq:fh_tof}
    N_{Tof} = (4r+1)(L^2 + \lfloor \log_2(L^2)\rfloor- w(L^2) +1).
\end{equation}
To get the T count, we first need to obtain the number of T gates needed to synthesize the $R_z$ gates. To do that, we split the rotation synthesis error $\Delta E_R$ into two parts such that
\begin{equation}
    \Delta E_R = s(1-y) \Delta E \rightarrow (x+z)(1-y) \Delta E,
\end{equation}
where we have let $s=x+z$. Then, we assign one part to synthesize the $R_z$ for the catalyst states, which leads to a T count of
\begin{equation}
   N_{T1}=(2\lfloor \log_2(L^2) \rfloor + 3) \Big(0.53 \log_2\Big(\frac{2\lfloor \log_2(L^2) \rfloor + 4}{z(1-y) \Delta E \cdot \tau}\Big) + 4.86 \Big),
\end{equation}
where we have used the repeat-until-success rotation synthesis from~\cite{Kliuchnikov2023shorterquantum}; the remaining part is assigned to synthesize the remaining $R_z$ gates, which require
\begin{equation}
    N_{T2}=(4r+1)\Big(0.53 \log_2\Big(\frac{4r+1}{x(1-y) \Delta E \cdot \tau}\Big) + 4.86 \Big)
\end{equation}
T gates. Including the directly applied T gates, we obtain a T count of
\begin{equation}
    N_T = 12rL^2 + N_{T1} + N_{T2}.
\end{equation}
Assuming 1 Toffoli = 2 T~\cite{Gidney2019efficientmagicstate}, the total Toffoli count is given by
\begin{equation}
   N_{Tot} = N_q \left (N_{Tof} + \frac{N_T}{2}\right ),
\end{equation}
where we remind the readers that $N_q \approx \frac{0.76 \pi}{y\tau \Delta E}$. 

As in the qubitization algorithm, we set $\Delta E = 0.51\% L^2$. Then, we numerically minimize $N_{Tot}$ over the parameters $x,y,z$ and $\tau$, under the constraint:
\begin{equation}
    \tau < \left( \frac{\sqrt{2}}{W}\right)^{1/3},
\end{equation}
such that $\delta < \sqrt{2}/r^2$, and we observe that $r\gtrsim 10$; thus, the constraint $\delta \ll 1$ in~\eqref{eq:trot_err} is satisfied. We find typically that $x \sim 0.01$, $\tau \sim 0.02$, $y\sim 0.6$, and $z \sim 0.001$ are optimal.

We additionally estimate the resource requirements for three other scenarios: (1) we use baseline HWP instead of catalyzed HWP, and we at most effect the same-angle $R_z$ gates in batches of $L^2/2$ using (2) catalyzed HWP and (3) baseline HWP. The batched approaches limit the number of ancilla qubits required. For all scenarios, we set $u=8$, $t=1$, and $\Delta E = 0.51\% L^2$.  All Toffoli count and qubit count, including ancilla qubits for QPE, RUS rotation synthesis and HWP, and catalyst qubis when necessary, are shown in Supplementary Table~\ref{tb:fh_trot_qre}.

\begin{table}[!ht]
\begin{tabular}{l|l|ll|ll|ll|ll}
    & & \multicolumn{2}{l|}{Catalyzed HWP}              & \multicolumn{2}{l|}{Baseline HWP}               & \multicolumn{2}{l|}{Batched, catalyzed HWP}     & \multicolumn{2}{l}{Batched, baseline HWP}       \\ \hline
$L$ & $W$ & \multicolumn{1}{l|}{$\#$ Toffoli} & $\#$ qubits & \multicolumn{1}{l|}{$\#$ Toffoli} & $\#$ qubits & \multicolumn{1}{l|}{$\#$ Toffoli} & $\#$ qubits & \multicolumn{1}{l|}{$\#$ Toffoli} & $\#$ qubits \\ \hline
4   & $2.82e2$ & \multicolumn{1}{l|}{$9.80e5$}     & 66          & \multicolumn{1}{l|}{$1.52e6$}     & 49          & \multicolumn{1}{l|}{$1.23e6$}     & 55          & \multicolumn{1}{l|}{$2.07e6$}     & 41          \\ \hline
6   & $6.54e2$ & \multicolumn{1}{l|}{$8.92e5$}     & 128         & \multicolumn{1}{l|}{$1.19e6$}     & 108         & \multicolumn{1}{l|}{$9.32e5$}     & 107         & \multicolumn{1}{l|}{$1.48e6$}     & 90          \\ \hline
8   & $1.16e3$ & \multicolumn{1}{l|}{$8.40e5$}     & 216         & \multicolumn{1}{l|}{$1.08e6$}     & 193         & \multicolumn{1}{l|}{$8.79e5$}     & 181         & \multicolumn{1}{l|}{$1.24e6$}     & 161         \\ \hline
10  & $1.83e3$ & \multicolumn{1}{l|}{$8.23e5$}     & 322         & \multicolumn{1}{l|}{$9.64e5$}     & 299         & \multicolumn{1}{l|}{$8.81e5$}     & 269         & \multicolumn{1}{l|}{$1.10e6$}     & 249         \\ \hline
12  & $2.63e3$& \multicolumn{1}{l|}{$8.16e5$}     & 458         & \multicolumn{1}{l|}{$9.20e5$}     & 432         & \multicolumn{1}{l|}{$8.86e5$}     & 383         & \multicolumn{1}{l|}{$1.05e6$}     & 360         \\ \hline
14  & $3.61e3$& \multicolumn{1}{l|}{$8.07e5$}     & 613         & \multicolumn{1}{l|}{$8.83e5$}     & 587         & \multicolumn{1}{l|}{$8.68e5$}     & 512         & \multicolumn{1}{l|}{$9.81e5$}     & 489         \\ \hline
16  &$4.69e3$ & \multicolumn{1}{l|}{$8.04e5$}     & 798         & \multicolumn{1}{l|}{$8.83e5$}     & 769         & \multicolumn{1}{l|}{$8.63e5$}     & 667         & \multicolumn{1}{l|}{$9.45e5$}     & 641         \\ \hline
18  & $5.82e3$& \multicolumn{1}{l|}{$8.02e5$}     & 1000         & \multicolumn{1}{l|}{$8.62e5$}     & 971         & \multicolumn{1}{l|}{$8.51e5$}     & 835         & \multicolumn{1}{l|}{$9.25e5$}     & 809         \\ \hline
20  & $7.21e3$& \multicolumn{1}{l|}{$8.00e5$}     & 1228        & \multicolumn{1}{l|}{$8.51e5$}     & 1199        & \multicolumn{1}{l|}{$8.54e5$}     & 1025        & \multicolumn{1}{l|}{$9.05e5$}     & 999         \\ \hline
22  & $8.71e3$& \multicolumn{1}{l|}{$8.00e5$}     & 1478        & \multicolumn{1}{l|}{$8.44e5$}     & 1449        & \multicolumn{1}{l|}{$8.40e5$}     & 1233        & \multicolumn{1}{l|}{$8.89e5$}     & 1207        \\ \hline
24  & $1.04e4$& \multicolumn{1}{l|}{$7.94e5$}     & 1760        & \multicolumn{1}{l|}{$8.39e5$}     & 1728        & \multicolumn{1}{l|}{$8.48e5$}     & 1469        & \multicolumn{1}{l|}{$8.87e5$}     & 1440        \\ \hline
26  & $1.22e4$& \multicolumn{1}{l|}{$8.01e5$}     & 2058        & \multicolumn{1}{l|}{$8.42e5$}     & 2026        & \multicolumn{1}{l|}{$8.46e5$}     & 1717        & \multicolumn{1}{l|}{$8.85e5$}     & 1688        \\ \hline
28  & $1.42e4$& \multicolumn{1}{l|}{$7.96e5$}     & 2383        & \multicolumn{1}{l|}{$8.36e5$}     & 2351        & \multicolumn{1}{l|}{$8.49e5$}     & 1988        & \multicolumn{1}{l|}{$8.77e5$}     & 1959        \\ \hline
30  & $1.63e4$& \multicolumn{1}{l|}{$7.98e5$}     & 2730        & \multicolumn{1}{l|}{$8.37e5$}     & 2698        & \multicolumn{1}{l|}{$8.41e5$}     & 2277        & \multicolumn{1}{l|}{$8.78e5$}     & 2248        \\ \hline
32  & $1.86e4$& \multicolumn{1}{l|}{$8.01e5$}     & 3108        & \multicolumn{1}{l|}{$8.42e5$}     & 3073        & \multicolumn{1}{l|}{$8.45e5$}     & 2593        & \multicolumn{1}{l|}{$8.86e5$}     & 2561       
\end{tabular}
\caption{Resource estimates of Trotter simulations and the Trotter error bounds $W$ of the $L\times L$ Fermi-Hubbard model with $u=8$, $t=1$, and $\Delta E = 0.51\% L^2$. Note that (i) catalyzed HWP always yields the lowest Toffoli count, (ii) batched, baseline HWP always yields the lowest qubit count, and (iii) batched, catalyzed HWP always yields similar Toffoli counts to baseline HWP, while using fewer qubits, except for when $L=4$.}
\label{tb:fh_trot_qre}
\end{table}

\subsection{Single-orbital Cuprate model}

\noindent\textbf{Trotter scheme and error:} As with the Fermi-Hubbard case, we consider an $L\times L$ lattice with periodic boundary conditions, i.e., $\Lambda=(\mathbbm{Z}_L)^2$, except that, in this case, we consider only values of $L$ that are multiples of four. We defined a new sub-lattice, $\Lambda' = ((4\mathbbm{Z})_L)^2 = \{ (0,0),(0,4),(4,0),(4,4),...,(L-4,L-4) \}$. With the coordinates $p_k$ defined as before, we now express the Hermitian summands in the decomposition of the Hamiltonian. First, the nn terms are:
\begin{equation}
    H_{nn,1} = t\sum_{i\in\Lambda'} \sum_{\sigma} \sum_{l,m=0}^1 \sum_{k=1}^4 h_{i+p_k+(2l,2m), i+p_{k+1}+(2l,2m),\sigma},
\end{equation}
and $H_{nn,2}$ is obtained by shifting $H_{nn,1}$ by $(-1,-1)$ on the lattice.
Second, the nnn terms are:
\begin{gather}
    H_{nnn,1}= t'\sum_{i\in\Lambda'} \sum_{\sigma} \sum_{l,m=0}^1 \sum_{k=1}^2 h_{i+p_k+(2l,2m), i+p_{k+2}+(2l,2m),\sigma},\\
    H_{nnn,2} = t'\sum_{i\in\Lambda'} \sum_{\sigma} \sum_{l,m=0}^1 \sum_{k=1}^2 h_{i+p_k-(1+2l,1+2m), i+p_{k+2}-(1+2l,1+2m),\sigma},\\
    H_{nnn,3} = t'\sum_{i\in\Lambda'} \sum_{\sigma} \sum_{l,m=0}^1 \sum_{k=1}^2 h_{i+p_k+(1+2l,2m), i+p_{k+2}+(1+2l,2m),\sigma},\\
    H_{nnn,4}= t'\sum_{i\in\Lambda'} \sum_{\sigma} \sum_{l,m=0}^1 \sum_{k=1}^2 h_{i+p_k+(2l,2m-1), i+p_{k+2}+(2l,2m-1),\sigma}.
\end{gather}
Third, the nnnn terms are
\begin{gather}
    H_{nnnn,1}= t''\sum_{i \in \Lambda'}\sum_{\sigma}\sum_{k=1}^4 h_{i + 2p_k, i+2p_{k+1},\sigma} + h_{i + 2p_k+(0,1), i+2p_{k+1}+(0,1),\sigma} \nonumber \\
    +h_{i + 2p_k + (1,0), i+2p_{k+1}+(1,0),\sigma}+h_{i + 2p_k+(1,1), i+2p_{k+1}+(1,1),\sigma}, \\
    H_{nnnn,2}= t''\sum_{i \in \Lambda'}\sum_{\sigma}\sum_{k=1}^4 h_{i + 2p_k+(2,2), i+2p_{k+1}+(2,2),\sigma}+ h_{i + 2p_k+(2,3), i+2p_{k+1}+(2,3),\sigma} \nonumber \\
    +h_{i + 2p_k+(3,2), i+2p_{k+1}+(3,2),\sigma}+h_{i + 2p_k+(3,3), i+2p_{k+1}+(3,3),\sigma}.
\end{gather}
Last, the on-site Coulomb term is the same as in~\eqref{eq:FH_on}. We order these nine terms according to the ordered list:
\begin{equation}
    \{ H_i \}_{i=1}^9 = \{ H_{nn,1},H_{nn,2},H_{nnn,1},H_{nnn,2},H_{nnn,3},H_{nnn,4},H_{nnnn,1},H_{nnnn,2},H_c \}.
\end{equation}

We use the automated symbolic commutator evaluation software from~\cite{Schubert_github,schubert2023trotter} to evaluate the Trotter error; the software is designed to automate the evaluation of commutator bounds of Fermi-Hubbard-like models. Applied to the Fermi-Hubbard model, it produces bounds with similarly sized constants as the analytical bounds from~\cite{campbell2021early}. We briefly describe how the software works and refer readers to~\cite{Schubert_github,schubert2023trotter} for further details. The software admits Hermitian operators of the form $\sum_{\bi \in \Lambda'} A_{\bi}$, where $A_{\bi}$ are local operators composing of either hopping operators or number operators and $\Lambda'$ is a sublattice of $\Lambda$ on which $A_{\bi}$ is defined. Each commutator will be of the form
\begin{equation}\label{eq:AB}
    \left [  \sum_{\bi \in \Lambda'} A_{\bi}, \sum_{\bi' \in \Lambda'} B_{\bi'} \right ] = \sum_{\bi \in \Lambda'} \left[A_{\bi},  \sum_{\bj \in \Lambda'} B_{\bi+\bj} \right],
\end{equation}
where the equality is implied by the periodic boundary conditions and translational invariance of the two Hermitian operators~\cite{schubert2023trotter}. Note that $A_{\bi}$ or $B_{\bi}$ could be a commutator. The software enumerates the lattice vectors $\bj$ for which $A_{\bi}$ and $B_{\bi+\bj}$ overlap in support. Then, it evaluates the right hand side of~\eqref{eq:AB} symbolically; if it evaluates to (1) quadratic/free-fermionic operators, their norms are computed exactly; if it evaluates to (2) non-quadratic operators, they are partitioned into clusters supported on no more than 14 fermionic modes, i.e., $\bi \in \Lambda'$, then diagonalized exactly, and finally summed up using triangle inequality. We report below the computed commutator error bound:
\begin{gather}
W = L^2(0.5562|t^3| + 3.5166|t^2 t'| +1.0147|t^2  t''| + 1.2652|t^2  u|  + 5.9063|t t'^2|+ 6.6727|t  t'  t''| \nonumber  \\
+ 2.7246|t t' u| + 1.4832|t t''^2| + 2.4294|tt''u| + 0.2018|tu^2| + 4.2510|t'^3| +5.7898|t'^2 t''|\nonumber \\
+2.2182|t'^2 u| + 4.2787|t' t''^2| + 2.8980|t' t'' u|+0.3333|t' u^2|
+0.7688|t''^3| +1.3761|t''^2 u|+0.2369|t'' u^2|),
\end{gather}
where we remind readers that $\delta \leq \tau^3 W/r^2$ as per~\eqref{eq:trot_err2}.

\noindent\textbf{Circuit implementation:} The implementation of the on-site and nn hopping terms are the same way as in the Fermi-Hubbard model. In what follows, we describe how the nnn and nnnn hopping terms are implemented. We perform the same JW transformation as in the qubitization case.

\underline{NNN hopping term $e^{-i H_{nnn,i}\theta}$}: First, we reorder the fermions in $H_{nnn,i}$ such that the labels of pairwise interacting fermions differ by 1 as follows: Each $H_{nnn,i}$ consists of hopping operators acting on diagonal pairs of fermions on non-overlapping plaquettes. For each such plaquette, we swap the top right and bottom right fermions using the fermionic swap operation defined in~\eqref{eq:nonlocal_fswap}; there are $O(L^2)$ plaquettes and each swap consists of $O(L)$ nn fermionic swaps, so $O(L^3)$ nn fermionic swaps are needed. Then, under JW transformation, all hopping operators will be of the form $e^{-i\theta (XX+YY)/2}$, which are mapped to a layer to same-angle $R_z(\theta)$ gates using Clifford gates. Finally, the fermionic swaps are uncomputed.

\underline{NNNN hopping term $e^{-i H_{nnnn,i}\theta}$}: First, we map $H_{nnnn,i}$ to $H_{nn,i}$ using fermionic swaps as follows. We group the hopping operators in $H_{nnnn,i}$ such that each group acts on fermions on the corners of $3\times 3$ squares. Then, we divide the lattice into $4\times 4$ squares in a way that each square consists of four such $3\times 3$ squares; for example, an $8\times 8$ lattice can be divided into four such $4\times 4$ squares. Next, for each $4\times 4$ square, we swap the middle two sites along every side, and swap diagonally the corners of the central $2\times 2$ plaquette. So, in total, we have applied six non-local fermionic swaps, each consisting of $O(L)$ nn fermionic swaps. There are $O(L^2)$ such $4\times 4$ squares, which means $O(L^3)$ nn fermionic swaps are needed to map $H_{nnnn,i}$ to $H_{nn,i}$. Next, we apply the circuit implementation of the nn hopping terms, before uncomputing the fermionic swaps.

\noindent\textbf{Resource estimation:} The total gate count is given by the gate count per approximated evolution $U(\tau,r)$ multiplied by the number of queries $N_q$ to it. To construct $U(\tau,r)$, $e^{-i\theta H_c}$ is applied $r$ times, $e^{-i\theta H_{nn,i}}$'s are applied $7r+1$ times, and $e^{-i\theta H_{nnn,i}}$'s are applied $8r$ times where adjacent evolutions of the same term are merged; note that each term is implemented using catalyzed HWP, the costs of which are delineated in the main text (see ``Space-time trade-off in Hamming-weight phasing"), up to Clifford gates and two-site fermionic Fourier transforms. Each $e^{-i\theta H_c}$ costs 1 $R_z$ gate and $L^2 + \lfloor \log_2(L^2)\rfloor- w(L^2) +1$ Toffoli gates. Each $e^{-i\theta H_{nn,i}}$ costs 1 $R_z$ gate, $4L^2$ T gates and $L^2 + \lfloor \log_2(L^2)\rfloor- w(L^2) +1$ Toffoli gates, and $e^{-i\theta H_{nnn,i}}$ costs 1 $R_z$ gate and $2L^2 + \lfloor \log_2(2L^2)\rfloor- w(2L^2) +1$ Toffoli gates, not including the synthesis costs of the catalyst states. The required catalyst states are $\bigotimes_{i=0}^{\lfloor \log_2(L^2) \rfloor} R_z(2^i \frac{u \tau}{r}) \ket{+...+}$ for $e^{-i\frac{u \tau}{2r} H_c}$, $\bigotimes_{i=0}^{\lfloor \log_2(L^2) \rfloor + 1} R_z(2^i \frac{t \tau}{r}) \ket{+...+}$ for $e^{-i\frac{t \tau}{2r} H_{nn,i}}$ and $e^{-i\frac{t \tau}{r} H_{nn,1}}$, $\bigotimes_{i=0}^{\lfloor \log_2(2L^2) \rfloor} R_z(2^i \frac{t' \tau}{r}) \ket{+...+}$ for $e^{-i\frac{t' \tau}{2r} H_{nnn,i}}$, and $\bigotimes_{i=0}^{\lfloor \log_2(L^2) \rfloor} R_z(2^i \frac{t'' \tau}{r}) \ket{+...+}$ for $e^{-i\frac{t'' \tau}{2r} H_{nnnn,i}}$. These states cost $3\lfloor \log_2(L^2) \rfloor + \lfloor \log_2(2L^2) \rfloor + 5$ $R_z$ gates.

The Toffoli count is
\begin{equation}\label{eq:cuprate_tof}
    N_{Tof} = (8r+1)(L^2 + \lfloor \log_2(L^2)\rfloor- w(L^2) +1) + 8r(2L^2 + \lfloor \log_2(2L^2)\rfloor- w(2L^2) +1).
\end{equation}
To get the T count, we first need to obtain the number of T gates needed to synthesize the $R_z$ gates. To do that, we split the rotation synthesis error $\Delta E_R$ into two parts such that
\begin{equation}
    \Delta E_R = s(1-y) \Delta E \rightarrow (x+z)(1-y) \Delta E,
\end{equation}
where we have let $s=x+z$. Then, we assign one part to synthesize the $R_z$ for the catalyst states, which leads to a T count of
\begin{equation}
   N_{T1}=(3\lfloor \log_2(L^2) \rfloor + \lfloor \log_2(2L^2) \rfloor + 5) \Big(0.53 \log_2\Big(\frac{3\lfloor \log_2(L^2) \rfloor + \lfloor \log_2(2L^2) \rfloor + 5}{z(1-y) \Delta E \cdot \tau}\Big) + 4.86 \Big),
\end{equation}
where we have used the repeat-until-success rotation synthesis from~\cite{Kliuchnikov2023shorterquantum}; the remaining part is assigned to synthesize the remaining $R_z$ gates, which require
\begin{equation}
    N_{T2}=(16r+1)\Big(0.53 \log_2\Big(\frac{16r+1}{x(1-y) \Delta E \cdot \tau}\Big) + 4.86 \Big)
\end{equation}
T gates. Including the directly applied T gates, we obtain a T count of
\begin{equation}
    N_T = 4L^2(7r+1) + N_{T1} + N_{T2}.
\end{equation}
Assuming 1 Toffoli = 2 T, the total Toffoli count is given by
\begin{equation}
   N_{Tot} = N_q \left (N_{Tof} + \frac{N_T}{2}\right ),
\end{equation}
where we remind the readers that $N_q \approx \frac{0.76 \pi}{y\tau \Delta E}$. 

As in the qubitization algorithm, we set $\Delta E = 0.51\% L^2$. Then, we numerically minimize $N_{Tot}$ over the parameters $x,y,z$ and $\tau$, under the constraint:
\begin{equation}
    \tau < \left( \frac{\sqrt{2}}{W}\right)^{1/3},
\end{equation}
such that $\delta < \sqrt{2}/r^2$, and we observe that $r\gtrsim 10$; thus, the constraint $\delta \ll 1$ in~\eqref{eq:trot_err} is satisfied. We find typically that $x\sim 0.01$, $z \sim 0.05$, $y\sim 0.6$, and $0.006 \lesssim \tau \lesssim 0.1$ are optimal.

We additionally estimate the resource requirements for three other scenarios: (1) we use baseline HWP instead of catalyzed HWP, and we at most effect the same-angle $R_z$ gates in batches of $L^2/2$ using (2) catalyzed HWP and (3) baseline HWP. The batched approaches limit the number of ancilla qubits required. For all scenarios, we set $u=8$, $t=1$, $t'=0.3$, $t''=0.2$, and $\Delta E = 0.51\% L^2$, as in the qubitization case. All Toffoli count and qubit count, including ancilla qubits for QPE, RUS rotation synthesis and HWP, and catalyst qubis when necessary, are shown in Supplementary Table~\ref{tb:cu_trot_qre}.

The Toffoli count of the catalyzed HWP implementation is roughly an order of magnitude higher than that of the FH simulation at the same lattice size, which can be explained as follows: For an $L \times L$ cuprate simulation, the Toffoli count, at leading order, is $24 rL^2$ according to~\eqref{eq:cuprate_tof}, where $r$ scales as $\sqrt{W}$ according to~\eqref{eq:rW} and the factor of $24 L^2$ is due to the Trotter step circuit implementation. For an $L \times L$ FH simulation, the Toffoli count, at leading order, is $4 rL^2$ according to~\eqref{eq:fh_tof}. The values of $\sqrt{W}$ for the cuprate model, at the considered parameter setting, range from roughly 28 to 225 for $L=4$ to 32, which are roughly 1.6 times higher that $\sqrt{W}$ for the FH model, at the considered parameter setting; along with the factor of 6 difference between the costs of a Trotter step between the two models, we obtain roughly the order of magnitude difference between the Toffoli counts of FH and cuprate simulations. Furthermore, we observe that the Trotter step implementation has a $\sim 3$ times higher impact on the Toffoli count than the Trotter error bound. A similar analysis holds for other less Toffoli-efficient Trotter step implementations, e.g., baseline HWP, which will increase the Trotter step implementation but not affect the Trotter error.

\begin{table}[!ht]
\begin{tabular}{l|l|ll|ll|ll|ll}
    & & \multicolumn{2}{l|}{Catalyzed HWP}              & \multicolumn{2}{l|}{Baseline HWP}               & \multicolumn{2}{l|}{Batched, catalyzed HWP}     & \multicolumn{2}{l}{Batched, baseline HWP}       \\ \hline
$L$ & $W$& \multicolumn{1}{l|}{$\#$ Toffoli} & $\#$ qubits & \multicolumn{1}{l|}{$\#$ Toffoli} & $\#$ qubits & \multicolumn{1}{l|}{$\#$ Toffoli} & $\#$ qubits & \multicolumn{1}{l|}{$\#$ Toffoli} & $\#$ qubits \\ \hline
4   & $7.91e2$& \multicolumn{1}{l|}{$6.38e6$}     & 93          & \multicolumn{1}{l|}{$1.19e7$}     & 65          & \multicolumn{1}{l|}{$9.08e6$}     & 62          & \multicolumn{1}{l|}{$1.94e7$}     & 41          \\ \hline
8   & $3.16e3$&\multicolumn{1}{l|}{$5.25e6$}     & 295         & \multicolumn{1}{l|}{$7.08e6$}     & 257         & \multicolumn{1}{l|}{$6.07e6$}     & 192         & \multicolumn{1}{l|}{$1.01e7$}     & 161         \\ \hline
12  & $7.11e3$&\multicolumn{1}{l|}{$4.91e6$}     & 619         & \multicolumn{1}{l|}{$6.02e6$}     & 576         & \multicolumn{1}{l|}{$5.30e6$}     & 396         & \multicolumn{1}{l|}{$7.55e6$}     & 360         \\ \hline
16  & $1.26e4$&\multicolumn{1}{l|}{$5.22e6$}     & 1073        & \multicolumn{1}{l|}{$5.65e6$}     & 1025        & \multicolumn{1}{l|}{$5.45e6$}     & 682         & \multicolumn{1}{l|}{$6.40e6$}     & 641         \\ \hline
20  & $1.98e4$& \multicolumn{1}{l|}{$5.12e6$}     & 1647        & \multicolumn{1}{l|}{$5.40e6$}     & 1599        & \multicolumn{1}{l|}{$5.28e6$}     & 1040        & \multicolumn{1}{l|}{$6.00e6$}     & 999         \\ \hline
24  & $2.85e4$&\multicolumn{1}{l|}{$5.11e6$}     & 2357        & \multicolumn{1}{l|}{$5.39e6$}     & 2304        & \multicolumn{1}{l|}{$5.23e6$}     & 1486        & \multicolumn{1}{l|}{$5.87e6$}     & 1440        \\ \hline
28  & $3.87e4$& \multicolumn{1}{l|}{$5.09e6$}     & 3188        & \multicolumn{1}{l|}{$5.25e6$}     & 3135        & \multicolumn{1}{l|}{$5.18e6$}     & 2005        & \multicolumn{1}{l|}{$5.59e6$}     & 1959        \\ \hline
32  & $5.06e4$&\multicolumn{1}{l|}{$5.07e6$}     & 4155        & \multicolumn{1}{l|}{$5.21e6$}     & 4097        & \multicolumn{1}{l|}{$5.18e6$}     & 2612        & \multicolumn{1}{l|}{$5.51e6$}     & 2561       
\end{tabular}
\caption{Resource estimates of Trotter simulations and the Trotter error bounds $W$ of the $L\times L$ cuprate model with $u=8$, $t=1$, $t'=0.3$, $t''=0.2$, and $\Delta E = 0.51\% L^2$, as well as the Trotter error bounds $W$ at this parameter setting. Note that (i) catalyzed HWP always yields the lowest Toffoli count, (ii) batched, baseline HWP always yields the lowest qubit count, and (iii) batched, catalyzed HWP always yields similar Toffoli counts to baseline HWP, while using fewer qubits.}
\label{tb:cu_trot_qre}
\end{table}

\subsection{Two-orbital Pnictide model}

\noindent\textbf{Trotter scheme and error:} Once again, we consider an $L\times L$ lattice with periodic boundary conditions and even values of $L$. We will use the software from~\cite{schubert2023trotter,Schubert_github} to evaluate the Trotter error bounds. However, the software does not explicitly handle multi-orbital models. As a workaround, we embed a $(\mathbbm{Z}/L)^2$ lattice with two orbitals per site onto a $\Lambda = (\mathbbm{Z}/(2L))^2$ lattice, where the sites with only even and odd coordinates represent the two different orbitals, respectively, while the sites with mixed-parity coordinates are empty. Note that this lattice embedding is only to facilitate the evaluation of the commutator error bound; the JW transformation is still performed over the original lattice. We define a sub-lattice $\Lambda' = ((4\mathbbm{Z})/2L)^2$, and label $p_1 = (0,0)$, $p_2 = (1,0)$, $p_3=(1,1)$ and $p_4=(0,1)$. We decompose the Hamiltonian into fourteen terms, as follows:
\begin{gather}
    H_{nn,1} = t_1 \sum_{i\in\Lambda'} \sum_\sigma (h_{i+p_1, i+2p_4,\sigma} + h_{i+2p_2, i+2p_3,\sigma} + h_{i+p_3, i+p_3 + 2p_2,\sigma} + h_{i+p_3+2p_4, i+3p_3,\sigma}), \\
    H_{nn,3} = t_2 \sum_{i\in\Lambda'} \sum_\sigma (h_{i+p_1, i+2p_2,\sigma} + h_{i+2p_4, i+2p_3,\sigma} + h_{i+p_3, i+p_3 + 2p_4,\sigma} + h_{i+p_3+2p_2, i+3p_3,\sigma}), \\
    H_{nnn,1} = t_3 \sum_{i\in\Lambda'} \sum_\sigma (h_{i+p_1, i+2p_3,\sigma} + h_{i+2p_4, i+2p_2,\sigma} + h_{i+p_3, i+3p_3,\sigma} + h_{i+p_3+2p_4, i+p_3+2p_2,\sigma}), \\
    H_{nnn,5} = t_4 \sum_{i\in\Lambda'} \sum_\sigma (h_{i+p_1, i+3p_3,\sigma} + h_{i+p_3, i+2p_3,\sigma} - h_{i+2p_4, i+p_3+2p_2,\sigma} - h_{i+p_3+2p_4, i+2p_2,\sigma}), \\
    H_{c,1} = u \sum_{i\in\Lambda'} \sum_{k=1}^4 \left(n_{i+2p_k, \uparrow}-\frac{1}{2}\right) \left(n_{i+2p_k, \downarrow}-\frac{1}{2}\right) + \left(n_{i+2p_k + p_3, \uparrow}-\frac{1}{2}\right) \left(n_{i+2p_k + p_3, \downarrow}-\frac{1}{2}\right), \\
    H_{c,2} = v \sum_{i\in\Lambda'} \sum_{\sigma, \sigma'} \sum_{k=1}^4 \left(n_{i+2p_k, \sigma}-\frac{1}{2}\right) \left(n_{i+2p_k + p_3, \sigma'}-\frac{1}{2}\right)
\end{gather}
$H_{nn,2}$ is obtained from $H_{nn,1}$ by shifting all its lattice indices by $(2,2)$. $H_{nn,4}$ is obtained from $H_{nn,3}$ by shifting all its lattice indices by $(2,2)$. $H_{nnn,2}$, $H_{nnn,3}$, and $H_{nnn,4}$ are obtained from $H_{nnn,1}$ by shifting all its lattice indices by $(0,2)$, $(2,0)$, and $(2,2)$, respectively. $H_{nnn,6}$, $H_{nnn,7}$, and $H_{nnn,8}$ are obtained from $H_{nnn,5}$ by shifting all its lattice indices by $(0,2)$, $(2,0)$, and $(2,2)$, respectively.

We order these terms according to the following ordered list:
\begin{equation}
    \{ H_i\}_{i=1}^{14} = \{ H_{nn,1},H_{nn,2},H_{nn,3},H_{nn,4},H_{nnn,1},H_{nnn,2},...,H_{nnn,8},H_{c,1},H_{c,2} \}.
\end{equation}
We report below the computed commutator error bound:
\begin{gather}
W = L^2(0.25|t_1^3| + 1.3333|t_1^2 t_3| + 1.4524|t_1^2 t_4| + 0.3333|t_1^2 u| + 0.7233|t_1^2 v|
+0.6667|t_1 t_2 t_3| + 1.9374|t_1 t_2 t_4| \nonumber \\ 
+ 0.3398|t_1 t_2 u| + 1.037|t_1 t_2 v|+2.6667|t_1 t_3^2| + 5.6918|t_1 t_3 t_4| + 1.015|t_1 t_3 u| +2.6200|t_1 t_3 v| \nonumber \\
+ 4.2562|t_1 t_4^2| + 1.1301|t_1 t_4 u| + 2.8315|t_1 t_4 v| + 0.0833|t_1 u^2|
+0.2506|t_1 u v| + 3.8354|t_1 v^2| \nonumber \\
+ 0.25|t_2^3| + 1.3333|t_2^2 t_3|+ 1.4524|t_2^2 t_4| +0.3333|t_2^2 u| + 0.7363|t_2^2 v| +2.6667|t_2 t_3^2| + 5.6918|t_2 t_3 t_4| \nonumber \\
+1.0151|t_2 t_3 u| + 2.5783|t_2 t_3 v| +4.2562|t_2 t_4^2| +1.1279|t_2 t_4 u|
+ 2.7618|t_2 t_4 v| + 0.0833|t_2 u^2| \nonumber \\
+ 0.2506|t_2 u v| + 0.2397|t_2 v^2| + 2.8333|t_3^3| + 8.0|t_3^2 t_4| + 1.3333|t_3^2 u| + 2.7211|t_3^2 v| + 8.0|t_3 t_4^2| \nonumber \\
+ 2.3333|t_3 t_4 u| +4.3035|t_3 t_4 v| +0.1667|t_3 u^2| + 0.4714|t_3 u v|
+ 0.4714|t_3 v^2| +2.8333|t_4^3| \nonumber \\ 
+ 1.3333|t_4^2 u| + 2.7135|t_4^2 v| + 0.1667|t_4 u^2| +0.4714|t_4 u v| + 0.4714|t_4 v^2|).
\end{gather}

\noindent\textbf{Circuit implementation:} We perform the same JW transformation as in the qubitization case. 

\underline{On-site terms $e^{-i H_{c,i} \theta}$}: The intra-orbital on-site term $e^{-i H_{c,1} \theta}$ is implemented in the same way as the on-site term in the Fermi-Hubbard model. We implement inter-orbital on-site terms $e^{-i H_{c,2} \theta}$ in two batches. The first batch consists of terms acting on different spins, and the second batch consists of terms acting on the same spins. Both batches contain only $ZZ$ operators, whose evolution is then implemented in the same way as the intra-orbital terms, i.e., catalyzed HWP plus Cliffords.

\underline{NN hopping terms $e^{-i H_{nn,i}\theta}$}: The implementation for nn hopping terms from~\cite{campbell2021early} cannot be applied here because the nn hopping operators, which act on a given orbital $d \in \{x,y\}$, around a plaquette have different strengths, i.e., $t_1$ and $t_2$. Thus, we follow a strategy similar to the one used to implement nnn hopping terms in the cuprate model: apply fermionic swaps to reduce $H_{nn,i}$ into a sum of fermion pairs with labels differing by 1. Given a $H_{nn,i}$, we consider the four hopping operators of the same strength that act on a plaquette. We divide the four hopping operators into two pairs, which are distinguished by the direction in which the edges, whose ends are occupied by the fermions acted on by the operators, point in; they either point along the JW path or not; if not, then we pick an arbitrary fermion on the plaquette and swap it with the one diagonally across, using $O(L)$ nn fermionic swaps. There are $O(L^2)$ plaquettes, so we need $O(L^3)$ nn fermionic swaps in total. This maps $e^{-i H_{nn,i}\theta}$ into a product of $e^{-i (XX+YY)\theta}$ acting on distinct nn fermion pairs along the JW path; we then transform the $e^{-i (XX+YY)\theta}$ operators into a layer of same-angle $R_z$ gates, which we implement using catalyzed HWP.

\underline{NNN hopping terms $e^{-i H_{nnn,i}\theta}$}: Once again, adopt the same strategy as in the implementation of nnn hopping terms in the cuprate model. We consider a $H_{nnn,i}$ for a fixed $i$ and a plaquette. We focus first on the two hopping operators acting on diagonal pairs of the same type of orbital. Then, we use fermionic swaps, each consisting of $O(L^3)$ nn fermionic swaps, to `unwind' the diagonal hopping operators into nn hopping operators along the JW path, as we have done for the cuprate nnn hopping terms. We proceed to focus on the two hopping operators acting on diagonal pairs of different types of orbital. The difference here is that the fermionic swaps, which are used to `unwind' the diagonal hopping operators into nn hopping operators along the JW path, will each consist of $O(L^4)$ nn fermionic swaps. This is because the JW path is partitioned into two halves, one for each type of orbital and of length $O(L^2)$, and thus, swapping any pair of fermions of different orbital types within a plaquette will require $O(L^2)$ nn fermionic swaps. Dealing with all $O(L^2)$ plaquettes will then require $O(L^4)$ nn fermionic swaps. Once each $e^{-i H_{nnn,i}\theta}$ is boiled down to nn hopping operators, i.e., $e^{-i (XX+YY)\theta}$, along the JW path, we use Clifford gates to map them to a layer of same-angle $R_z(2\theta)$ gates that are implemented using catalyzed HWP.

\noindent\textbf{Resource estimation:} The total gate count is given by the gate count per approximated evolution $U(\tau,r)$ multiplied by the number of queries $N_q$ to it. To construct $U(\tau,r)$, $e^{-i\theta H_{c,1}}$ and $e^{-i\theta H_{c,2}}$ are applied $r$ times, $e^{-i\theta H_{nn,i}}$'s are applied $7r+1$ times, and $e^{-i\theta H_{nnn,i}}$'s are applied $8r$ times, where adjacent evolutions of the same term are merged; note that each term is implemented using catalyzed HWP, the costs of which are delineated in the main text (see ``Space-time trade-off in Hamming-weight phasing"), up to Clifford gates. Each $e^{-i\theta H_{c,1}}$ costs 1 $R_z$ gate and $2L^2 + \lfloor \log_2(2L^2)\rfloor- w(2L^2) +1$ Toffoli gates. Each $e^{-i\theta H_{c,2}}$ costs 2 $R_z$ gates and $2(2L^2 + \lfloor \log_2(2L^2)\rfloor- w(2L^2) +1)$ Toffoli gates. Each $e^{-i\theta H_{nn,i}}$ costs 1 $R_z$ gate and $4L^2 + \lfloor \log_2(4L^2)\rfloor- w(4L^2) +1$ Toffoli gates, and $e^{-i\theta H_{nnn,i}}$ costs 1 $R_z$ gate and $2L^2 + \lfloor \log_2(2L^2)\rfloor- w(2L^2) +1$ Toffoli gates, not including the synthesis costs of the catalyst states. The required catalyst states are $\bigotimes_{i=0}^{\lfloor \log_2(2L^2) \rfloor} R_z(2^i \frac{u \tau}{2r}) \ket{+...+}$ for $e^{-i\frac{u \tau}{4r} H_{c,1}}$, $\bigotimes_{i=0}^{\lfloor \log_2(2L^2) \rfloor} R_z(2^i \frac{v \tau}{2r}) \ket{+...+}$ for $e^{-i\frac{v \tau}{4r} H_{c,2}}$, $\bigotimes_{i=0}^{\lfloor \log_2(4L^2) \rfloor + 1} R_z(2^i \frac{t_{1} \tau}{r}) \ket{+...+}$ for $e^{-i\frac{t_{1} \tau}{2r} H_{nn,1}}$ and $e^{-i\frac{t_1 \tau}{r} H_{nn,1}}$, $\bigotimes_{i=0}^{\lfloor \log_2(4L^2) \rfloor} R_z(2^i \frac{t_{2} \tau}{r}) \ket{+...+}$ for $e^{-i\frac{t_{2} \tau}{2r} H_{nn,2}}$, $\bigotimes_{i=0}^{\lfloor \log_2(2L^2) \rfloor} R_z(2^i \frac{t_{3} \tau}{r}) \ket{+...+}$ for $e^{-i\frac{t_{3} \tau}{2r} H_{nnn,3}}$, and $\bigotimes_{i=0}^{\lfloor \log_2(2L^2) \rfloor} R_z(2^i \frac{t_{4} \tau}{r})\ket{+...+}$ for $e^{-i\frac{t_{4} \tau}{2r} H_{nnn,4}}$. These states cost $2 \lfloor \log_2(4L^2)\rfloor + 4 \lfloor \log_2(2L^2)\rfloor + 7$ $R_z$ gates.

The Toffoli count is
\begin{equation}\label{eq:pn_tof}
    N_{Tof} = (7r+1)(4L^2 + \lfloor \log_2(4L^2)\rfloor- w(4L^2) +1) + 11r(2L^2 + \lfloor \log_2(2L^2)\rfloor- w(2L^2) +1).
\end{equation}
To get the T count, we first need to obtain the number of T gates needed to synthesize the $R_z$ gates. To do that, we split the rotation synthesis error $\Delta E_R$ into two parts such that
\begin{equation}
    \Delta E_R = s(1-y) \Delta E \rightarrow (x+z)(1-y) \Delta E,
\end{equation}
where we have let $s=x+z$. Then, we assign one part to synthesize the $R_z$ for the catalyst states, which leads to a T count of
\begin{equation}
   N_{T1}=(2 \lfloor \log_2(4L^2)\rfloor + 4 \lfloor \log_2(2L^2)\rfloor + 7) \Big(0.53 \log_2\Big(\frac{2 \lfloor \log_2(4L^2)\rfloor + 4 \lfloor \log_2(2L^2)\rfloor + 7}{z(1-y) \Delta E \cdot \tau}\Big) + 4.86 \Big),
\end{equation}
where we have used the repeat-until-success rotation synthesis from~\cite{Kliuchnikov2023shorterquantum}; the remaining part is assigned to synthesize the remaining $R_z$ gates, which require
\begin{equation}
    N_{T2}=(18r+1)\Big(0.53 \log_2\Big(\frac{18r+1}{x(1-y) \Delta E \cdot \tau}\Big) + 4.86 \Big)
\end{equation}
T gates. Including the directly applied T gates, we obtain a T count of
\begin{equation}
    N_T = N_{T1} + N_{T2}.
\end{equation}
Assuming 1 Toffoli = 2 T, the total Toffoli count is given by
\begin{equation}
   N_{Tot} = N_q \left (N_{Tof} + \frac{N_T}{2}\right ),
\end{equation}
where we remind the readers that $N_q \approx \frac{0.76 \pi}{y\tau \Delta E}$. 

As in the qubitization algorithm, we set $\Delta E = 0.51\% L^2$. Then, we numerically minimize $N_{Tot}$ over the parameters $x,y,z$ and $\tau$, under the constraint:
\begin{equation}
    \tau < \left( \frac{\sqrt{2}}{W}\right)^{1/3} ,
\end{equation}
such that $\delta < \sqrt{2}/r^2$, and we observe that $r\gtrsim 10$; thus, the constraint $\delta \ll 1$ in~\eqref{eq:trot_err} is satisfied. We find typically that $0.01 \lesssim x \lesssim 0.03$, $0.1 \lesssim \tau \lesssim 0.2$, $y\sim 0.6$, and $z\sim 0.005$ are optimal.

We additionally estimate the resource requirements for three other scenarios: (1) we use baseline HWP instead of catalyzed HWP, and we at most effect the same-angle $R_z$ gates in batches of $L^2/2$ using (2) catalyzed HWP and (3) baseline HWP. The batched approaches limit the number of ancilla qubits required. For all scenarios, we set $u=8$, $t_1=1$, $t_2=0.85$, $t_3=0.85$, $t_4=0.85$, and $\Delta E = 0.51\% L^2$, as in the qubitization case.  All Toffoli count and qubit count, including ancilla qubits for QPE, RUS rotation synthesis and HWP, and catalyst qubis when necessary, are shown in Supplementary Table~\ref{tb:pn_trot_qre}.

The Toffoli count of the catalyzed HWP implementation is roughly 70 times higher than that of the FH simulation at the same lattice size, which can be explained as follows: For an $L \times L$ pnictide simulation, the Toffoli count, at leading order, is $50 rL^2$ according to~\eqref{eq:pn_tof}, where $r$ scales as $\sqrt{W}$ according to~\eqref{eq:rW} and the factor of $50 L^2$ is due to the Trotter step circuit implementation. For an $L \times L$ FH simulation, the Toffoli count, at leading order, is $4 rL^2$ according to~\eqref{eq:fh_tof}. The values of $\sqrt{W}$ for the pnictide model, at the considered parameter setting, range from roughly 106 to 850 for $L=4$ to 32, which are roughly 6 times higher that $\sqrt{W}$ for the FH model, at the considered parameter setting; along with the factor of 12.5 difference between the costs of a Trotter step between the two models, we obtain roughly the factor of 70 difference between the Toffoli counts of FH and pnictide simulations. Furthermore, we observe that the Trotter step implementation has a $\sim 2$ times higher impact on the Toffoli count than the Trotter error bound. A similar analysis holds for other less Toffoli-efficient Trotter step implementations, e.g., baseline HWP, which will increase the Trotter step implementation but not affect the Trotter error.

\begin{table}[!ht]
\begin{tabular}{l|l|ll|ll|ll|ll}
    & & \multicolumn{2}{l|}{Catalyzed HWP}              & \multicolumn{2}{l|}{Baseline HWP}               & \multicolumn{2}{l|}{Batched, catalyzed HWP}     & \multicolumn{2}{l}{Batched, baseline HWP}       \\ \hline
$L$ & $W$& \multicolumn{1}{l|}{$\#$ Toffoli} & $\#$ qubits & \multicolumn{1}{l|}{$\#$ Toffoli} & $\#$ qubits & \multicolumn{1}{l|}{$\#$ Toffoli} & $\#$ qubits & \multicolumn{1}{l|}{$\#$ Toffoli} & $\#$ qubits \\ \hline
4   &$1.14e4$& \multicolumn{1}{l|}{$4.16e7$}     & 175          & \multicolumn{1}{l|}{$7.59e7$}     & 129         & \multicolumn{1}{l|}{$8.35e7$}     & 101          & \multicolumn{1}{l|}{$1.86e8$}     & 73          \\ \hline
6   &$2.56e4$& \multicolumn{1}{l|}{$3.57e7$}     & 341         & \multicolumn{1}{l|}{$5.36e7$}     & 288         & \multicolumn{1}{l|}{$5.31e7$}     & 197         & \multicolumn{1}{l|}{$1.14e8$}     & 162          \\ \hline
8   &$4.54e4$& \multicolumn{1}{l|}{$3.36e7$}     & 573         & \multicolumn{1}{l|}{$4.53e7$}     & 513         & \multicolumn{1}{l|}{$4.57e7$}     & 331         & \multicolumn{1}{l|}{$8.76e7$}     & 289         \\ \hline
10  &$7.10e4$& \multicolumn{1}{l|}{$3.28e7$}     & 859         & \multicolumn{1}{l|}{$3.97e7$}     & 799         & \multicolumn{1}{l|}{$3.99e7$}     & 491         & \multicolumn{1}{l|}{$6.50e7$}     & 449         \\ \hline
12  &$1.02e5$& \multicolumn{1}{l|}{$3.25e7$}     & 1219         & \multicolumn{1}{l|}{$3.78e7$}     & 1152         & \multicolumn{1}{l|}{$3.83e7$}     & 697         & \multicolumn{1}{l|}{$5.92e7$}     & 648         \\ \hline
14  &$1.39e5$& \multicolumn{1}{l|}{$3.31e7$}     & 1634         & \multicolumn{1}{l|}{$3.58e7$}     & 1567       & \multicolumn{1}{l|}{$3.63e7$}     & 930         & \multicolumn{1}{l|}{$5.09e7$}     & 881         \\ \hline
16  &$1.82e5$& \multicolumn{1}{l|}{$3.13e7$}     & 2123         & \multicolumn{1}{l|}{$3.56e7$}     & 2049       & \multicolumn{1}{l|}{$3.60e7$}     & 1209         & \multicolumn{1}{l|}{$4.95e7$}     & 1153         \\ \hline
18  &$2.30e5$& \multicolumn{1}{l|}{$3.11e7$}     & 2665         & \multicolumn{1}{l|}{$3.42e7$}     & 2591       & \multicolumn{1}{l|}{$3.48e7$}     & 1513        & \multicolumn{1}{l|}{$4.46e7$}     & 1457         \\ \hline
20  &$2.84e5$& \multicolumn{1}{l|}{$3.14e7$}     & 3273        & \multicolumn{1}{l|}{$3.35e7$}     & 3199        & \multicolumn{1}{l|}{$3.43e7$}     & 1855        & \multicolumn{1}{l|}{$4.20e7$}     & 1799         \\ \hline
22  &$3.44e5$& \multicolumn{1}{l|}{$3.10e7$}     & 3943        & \multicolumn{1}{l|}{$3.31e7$}     & 3869        & \multicolumn{1}{l|}{$3.36e7$}     & 2231        & \multicolumn{1}{l|}{$3.98e7$}     & 2175        \\ \hline
24  &$4.09e5$& \multicolumn{1}{l|}{$3.10e7$}     & 4689        & \multicolumn{1}{l|}{$3.28e7$}     & 4608        & \multicolumn{1}{l|}{$3.38e7$}     & 2655        & \multicolumn{1}{l|}{$3.97e7$}     & 2592        \\ \hline
26  &$4.80e5$& \multicolumn{1}{l|}{$3.28e7$}     & 5487        & \multicolumn{1}{l|}{$3.27e7$}     & 5406        & \multicolumn{1}{l|}{$3.34e7$}     & 3103        & \multicolumn{1}{l|}{$3.82e7$}     & 3040        \\ \hline
28  &$5.56e5$& \multicolumn{1}{l|}{$3.11e7$}     & 6352        & \multicolumn{1}{l|}{$3.25e7$}     & 6271        & \multicolumn{1}{l|}{$3.33e7$}     & 3590        & \multicolumn{1}{l|}{$3.73e7$}     & 3527        \\ \hline
30  &$6.39e5$& \multicolumn{1}{l|}{$3.08e7$}     & 7279        & \multicolumn{1}{l|}{$3.25e7$}     & 7198        & \multicolumn{1}{l|}{$3.31e7$}     & 4111        & \multicolumn{1}{l|}{$3.64e7$}     & 4048        \\ \hline
32  &$7.27e5$& \multicolumn{1}{l|}{$3.08e7$}     & 8281        & \multicolumn{1}{l|}{$3.23e7$}     & 8193        & \multicolumn{1}{l|}{$3.32e7$}     & 4679        & \multicolumn{1}{l|}{$3.67e7$}     & 4609       
\end{tabular}
\caption{Resource estimates of Trotter simulations and the Trotter error bounds $W$ of the $L\times L$ pnictide model with $u=8$, $t_1=1$, $t_2=0.85$, $t_3=0.85$, $t_4=0.85$, and $\Delta E = 0.51\% L^2$, as well as the Trotter error bounds $W$,  at this parameter setting. Note that (i) catalyzed HWP always yields the lowest Toffoli count, (ii) batched, baseline HWP always yields the lowest qubit count, and (iii) batched, catalyzed HWP always yields similar Toffoli counts to baseline HWP, while using fewer qubits.}
\label{tb:pn_trot_qre}
\end{table}
\clearpage

\section{Hamming-Weight Phasing Circuits}
\begin{figure}[!ht]
\centering
\begin{adjustbox}{width=0.75\textwidth}
\begin{quantikz}
\lstick{$\ket{x_1}$}& \gate[3][1.25cm]{HA} & \\
\lstick{$\ket{x_2}$}& \gateoutput{$s$} & \\
\lstick{$\ket{0}$}& \gateoutput{$c$} & 
\end{quantikz}
=
\begin{quantikz}
\lstick{$\ket{x_1}$}& \ctrl{1} & \ctrl{1} & \rstick{$\ket{x_1}$}\\
\lstick{$\ket{x_2}$}& \control{} & \targ{} & \rstick{$\ket{x_1 \oplus x_2}$}\\
\setwiretype{n} & \wire[u]{q} & \setwiretype{q} & \rstick{$\ket{x_1 \land x_2}$}
\end{quantikz}
,
\begin{quantikz}
    \lstick{$\ket{x_1}$}& \gate[4][1.25cm]{FA} & \\
    \lstick{$\ket{x_2}$}& & \\
    \lstick{$\ket{x_3}$}& \gateoutput{$s$} & \\
    \lstick{$\ket{0}$}& \gateoutput{$c$} & 
\end{quantikz}
=
\begin{quantikz}
    \lstick{$\ket{x_1}$} & \ctrl{2}&  & \ctrl{3} & & \ctrl{2}&\rstick{$\ket{x_1}$}\\
    \lstick{$\ket{x_2}$} & \targ{}& \ctrl{1} & &\ctrl{1} & \targ{}&\rstick{$\ket{x_2}$}\\
    \lstick{$\ket{x_3}$} &\targ{} & \control{} & &\targ{} & \targ{}&\rstick{$\ket{x_1 \oplus x_2 \oplus x_3}$}\\
    \setwiretype{n} & & \wire[u]{q} & \targ{}\setwiretype{q}& & &\rstick{$\ket{MAJ(x_1,x_2, x_3)}$}
\end{quantikz}
\end{adjustbox}
,
\begin{adjustbox}{width=0.6\textwidth}
\begin{quantikz}[transparent]
    \lstick{$\ket{x_0}$} & \gate[4][1.25cm]{FA} & & &&&&&\\
    \lstick{$\ket{x_1}$} & & &&&&&&\\
    \lstick{$\ket{0}$} & \gateoutput{$c$}& &&&\gate[11][1.25cm]{FA}&&&\\
    \lstick{$\ket{x_2}$} & \gateoutput{$s$}&\gate[4][1.25cm]{FA}&&&\linethrough&&&\\
    \lstick{$\ket{x_3}$} & & &&&\linethrough&&&\\
    \lstick{$\ket{0}$} & &\gateoutput{$c$}&&&&& &\\
    \lstick{$\ket{x_4}$} & &\gateoutput{$s$}&\gate[4][1.25cm]{FA}&&\linethrough&&&\\
    \lstick{$\ket{x_5}$} & & &&&\linethrough&&&\\
    \lstick{$\ket{0}$} & &&\gateoutput{$c$}&&\gateoutput{s}&\gate[6, label style={yshift=-0.2cm}][1.25cm]{HA}& &\\
    \lstick{$\ket{x_6}$} & & &  \gateoutput{$s$}&\gate[3][1.25cm]{HA}&\linethrough& \linethrough&&\\
    \lstick{$\ket{0}$} & & &&\gateoutput{$c$}&\linethrough&\gateoutput{s}& &\rstick{$\ket{y_1}$}\\
    \lstick{$\ket{x_7}$} & & &&\gateoutput{$s$}& \linethrough&\linethrough& \linethrough&\rstick{$\ket{y_0}$}\\
    \lstick{$\ket{0}$} & & &&&\gateoutput{$c$}&\linethrough&\gate[3,label style={yshift=0cm}][1.25cm]{HA}\gateoutput{$c$}&\rstick{$\ket{y_3}$}\\
    \lstick{$\ket{0}$} & & &&&&\gateoutput{$c$}&&\\
    \lstick{$\ket{0}$} & & &&&& & \gateoutput{$s$}&\rstick{$\ket{y_2}$}
\end{quantikz}
\end{adjustbox}
\caption{Hamming-weight computation circuit for an 8-bit input. The HA and FA operations are half- and full-adder, respectively.}
\label{fig:Hamming_weight}
\end{figure}

\begin{figure}[!ht]
\centering
\includegraphics[width = \textwidth]{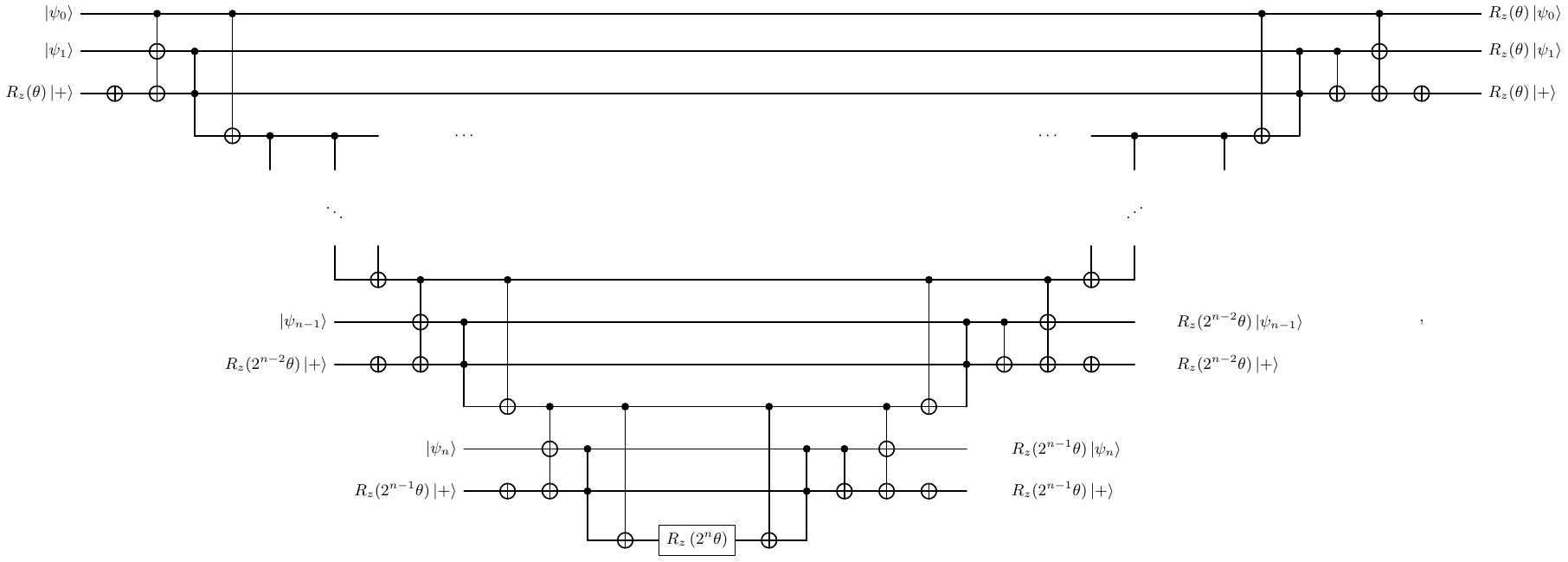}
\caption{Generalized phase-gradient circuit taken from equation 168 of~\cite{wang2024option}, which is derived from recursive applications of the phase-catalysis circuit in figure 14 of~\cite{Gidney2019efficientmagicstate}. Note that $\ket{\psi_0}$ and the gates controlled by it can be entirely removed for the catalyzed Hamming-weight phasing.}
\label{fig:phase_gadget}
\end{figure}

\nocite{}
\bibliographystyle{apsrev4-2}
\bibliography{refSM}